\documentclass[12pt,oneside]{amsart}
\usepackage[margin=1in]{geometry}
\usepackage{amsmath,amsthm,amssymb,amsfonts}
\usepackage{graphicx}
\usepackage{latexsym}
\usepackage[T2A]{fontenc}
\usepackage[utf8]{inputenc}
\usepackage[english]{babel}
\usepackage{natbib}
\usepackage{ragged2e}
\usepackage{setspace}
\usepackage[table]{xcolor}
\usepackage{csquotes}
\usepackage{ulem}
\usepackage{lineno}
\usepackage{thmtools}
\usepackage{hyperref}
\usepackage{diagbox}
\usepackage{multirow}
\usepackage{cleveref}
\usepackage{booktabs}

\usepackage{enumerate,tikz}
\usepackage{xcolor}

\newtheoremstyle{dotless}{}{}{\itshape}{}{\bfseries}{}{ }{}
\theoremstyle{dotless}
\newtheorem{lemma}{Lemma}
\newtheorem{theorem}{Theorem}
\newtheorem*{remark}{Remark}
\newtheorem{corollary}{Corollary}
\newtheorem{example}{Example}
\newtheorem{case}{Case}
\newcommand\xqed[1]{\leavevmode\unskip\penalty99999\hbox{}\nobreak\hfill\quad\hbox{#1}}
\newcommand\closeex{\xqed{$\triangle$}}

\newenvironment{property}[1]{\begin{justify}\textbf{#1}}{\end{justify}}

\begin{document}
\normalem

\thispagestyle{empty}

\title[]{Fair and consistent prize allocation in competitions}

\author{Bas J. Dietzenbacher}

\author{Aleksei Y. Kondratev}

\thanks{\phantom{ } b.dietzenbacher@maastrichtuniversity.nl --- Department of Quantitative Economics, Maastricht University, Maastricht, The Netherlands --- http://orcid.org/0000-0002-8023-0273\\
\phantom{ } akondratev@hse.ru --- corresponding author --- HSE University, International Laboratory of Game Theory and Decision Making, Russia; Institute for Regional Economic Studies RAS, St.~Petersburg, Russia --- http://orcid.org/0000-0002-8424-8198\\
\phantom{ } We are grateful to Alexander Nesterov, Egor Ianovski, Herv{\'e} Moulin, and other colleagues from the International Laboratory of Game Theory and Decision Making for helpful comments. Special thanks to William Thomson (University of Rochester) for detailed comments and suggestions. We thank David Connolly (HSE University) for help with language editing.\\
}


\begin{abstract}
Given the final ranking of a competition, how should the total prize endowment be allocated among the competitors? We study consistent prize allocation rules satisfying elementary solidarity and fairness principles. In particular, we axiomatically characterize several families of rules satisfying anonymity, order preservation, and endowment monotonicity, which all fall between the Equal Division rule and the Winner-Takes-All rule. Specific characterizations of rules and subfamilies are directly obtained. To our knowledge, this paper is the first that develops and motivates prize allocation rules for rank-order competitions directly from axioms.

\bigskip


\bigskip



\end{abstract}

\maketitle

\thispagestyle{empty}

\newpage

\thispagestyle{empty}

\section{Introduction}

Innovation and crowdsourcing competitions, sales competitions in companies and sporting events often take the form of rank-order tournaments \citep{KalraShi01,Szymanski03,TerwieschXu08,ArchakSundararajan09}. Each such tournament is held once, and the participants know the rules and the prize structure in advance. The absolute result of a participant can be determined by, for example, the volume of sales in a sales competition, the number of strokes in golf, or the time of elimination from a poker tournament. However, a feature of the ranking format is that participants receive prizes according to their relative results, while tournament organizers are free to choose the prize structure.

Examples of prize structures are presented in Table~\ref{table:golfpoker}. The PGA TOUR conducts many regular golf tournaments with different prize endowments, but the same rules for distributing prizes. Table~\ref{table:golfpoker} shows examples of two such tournaments with a prize endowment of \$9.3 million (The Genesis Invitational) and \$6.6 million (Safeway Open). The winner of the tournament receives 18\% of the total prize endowment, the runner-up 10.9\%, and so on. WCOOP Poker Tournament has a comparable prize endowment but follows a different prize-money distribution. What exactly are the similarities and differences of various prize structures? What general principles can tournament organizers follow when choosing a prize structure?

\begin{table}[ht!]
\caption{Examples of single rank-order tournaments}\label{table:golfpoker}
{\small
\begin{center}
\begin{tabular}{ccccccccccccc}
\multicolumn{13}{c}{(a) WCOOP 2019 Main Event poker tournament}\\
\toprule
Position&	1&	2&	3&	4&	5&	6&	7&	8&	9&	10&	Top 10&	Total\\
\hline
Prize&	1,666&	1,188&	847&	603&	430&	307&	219&	156&	111&	79&	5,605&	11,180\\
\hline
Share& 	14.9&	10.6&	7.57&	5.40&	3.85&	2.74&	1.96&	1.39&	0.99&	0.71&	50.1&	100\\
\bottomrule
\end{tabular}
\begin{tabular}{ccccccccccccc}
\\
\multicolumn{13}{c}{(b) PGA TOUR 2019/20 golf tournaments}\\
\toprule
Position&	1&	2&	3&	4&	5&	6&	7&	8&	9&	10&	Top 10&	Total\\
\hline
Genesis&	1,674&	1,014&	642&	456&	381&	337&	314&	291&	272&	253&	5,633&	9,300\\
Safeway&	1,188&	719&	455&	323&	271&	239&	223&	206&	193&	180&	3,998&	6,600\\
\hline
Share&	18.0&	10.9&	6.9&	4.9&	4.1&	3.63&	3.38&	3.13&	2.93&	2.73&	60.6&	100\\
\bottomrule
\end{tabular}
\end{center}}
\vspace{0.2cm}
\footnotesize{\emph{Notes}: the prize endowment, the total prize for top 10 positions, and the prize for each position from 1 to 10 are given in thousands of dollars. The shares are given in percentages of the total prize endowment.}
\end{table}

This paper examines prize structures in terms of fairness principles. Since the basic principles of justice must be universal and clear, they must also be found in particular examples of tournaments and competitions. Consistent with the principle of \emph{anonymity}, is that the participants’ rewards depend only on their position in the competition. That is, when the organizers award a prize for a position regardless of which participant takes that position. According to the principle of \emph{order preservation}, a higher position does not correspond to a lower reward. This creates the right incentives for participants, as the efforts made during preparation for and participation in the tournament help the competitor to end up higher in the final ranking and receive a more valuable reward. Thirdly, \emph{endowment monotonicity} is satisfied when an increase in the total prize fund does not decrease the reward for any position. This creates uniform incentives for all participants, as both leaders and outsiders are interested in increasing the total prize endowment of the tournament.\footnote{E.g., in some poker tournaments, players can re-buy and add-on chips, which increases the total prize money. Also, during some e-sports tournaments, viewers and sponsors can add to the total prize money.}

To distribute prizes, the organizers need to know three things: the list of the participants, the results of the competition, and the size of the prize endowment. For each such triple, the \emph{prize allocation rule} used should uniquely distribute the prize endowment among the participants. We call a rule for the distribution of prizes \emph{‘fair’} if it satisfies the principles of anonymity, order preservation, and endowment monotonicity. Anonymity and order preservation are standard principles for single rank-order tournaments; see the literature overview in subsection~\ref{sec:overview}. Endowment monotonicity is a standard principle in fair allocation literature; we refer to resource monotonicity in the book by \citet{Moulin03book}. These three principles of justice are so undemanding that even together they do not exclude any visible class of prize structures.

The universal principle of \emph{consistency} is often applied in fair allocation problems \citep{BalinskiYoung82,Thomson2012}. We apply this principle to the prize allocation problem as follows. Suppose that the participants have some ranking in a competition and receive the corresponding prizes. Then some participants leave with their prizes. The remaining participants decide to redistribute the remaining prize endowment, taking into account the modified composition of the participants. The rule is consistent if such a redistribution does not change the amount the remaining participants receive.

Consistent rules may help to stabilize the set of participants in a competition. Suppose a coalition of participants are unsatisfied with the prize money they are receiving. They could agree among themselves to leave the competition, and organize their own, smaller event, where they will divide the prize money in a manner they find more beneficial. However, if the organizer used a consistent rule, the participants would have earned the same amount in the original competition, and would have no incentive to leave.

Although consistency with respect to single rank-order tournaments is demanding, it is often observed in reality. For example, the Equal Division (ED) rule and the Winner-Takes-All (WTA) rule are consistent. As a more complex example, suppose that in a competition with 100 participants, the company awards a number of fixed-size cash prizes of \$2,000. The limited prize endowment provides a certain number of equal prizes to the participants with the highest positions, and any remaining balance goes to the next participant. So, if the prize fund is \$6,000, then the three best participants A, B, and C win \$2,000 each. If the fund grows to \$11,000, then the top five participants A, B, C, D, and E win \$2,000 each, and the participant in sixth position, F wins \$1,000.\footnote{Similarly, satellites of poker tournaments follow consistent prize structures.} Consistency requires that if the organizer has already transferred payments to participants B, C, and F, then re-applying the rule to the competition without participants B, C, and F and with \$6,000 prize endowment will leave everything as before: participants A, D, and E receive \$2,000 each, since they are the best among the remaining participants.

How do \emph{fair and consistent} prize distribution rules look like? The Equal Division rule, the Winner-Takes-All rule, and the aforementioned rule illustrate the main feature of such rules. Our first main result shows that all fair and consistent rules are some combination of these three rules and we call them \emph{interval rules} (Theorem~\ref{theorem:cons}). Since this is still a large family, we can strengthen our requirements for its properties.

In particular, we can strengthen order preservation. We say that the rule satisfies \emph{strict order preservation} if prizes for positions from the first to the last form a strictly decreasing sequence. Unfortunately, there are no fair and consistent rules for the distribution of prizes that satisfy strict order preservation. On the other hand, the Winner-Takes-All rule is the only fair and consistent rule in which the prize for the first position always exceeds the prize for the last position (Corollary~\ref{cor:sop}). 

We can also strengthen endowment monotonicity. To do this, we examine three more stringent properties. The first strengthening of endowment monotonicity is \emph{strict endowment monotonicity}. This property requires that with an increase in the prize endowment, the prize for each position strictly increases. The Equal Division rule is the only fair and consistent rule that satisfies strict endowment monotonicity (Corollary~\ref{cor:ED}).

The second strengthening is \emph{winner strict endowment monotonicity}. This property in addition to endowment monotonicity requires that with an increase in the prize endowment, the prize for the first position strictly increases. This leads to the following subfamily of rules. Each such rule sets a maximum size of an individual prize, which all competitors receive equally regardless of their position, while all the excess goes to the winner (Corollary~\ref{cor:WTS}). Therefore, we call this a \emph{Winner-Takes-Surplus (WTS) rule}. For example, the size of a laboratory’s premium fund is often recognized only at the end of the year. The head of the laboratory with 10 employees can consider a fair premium of \$2,000. Then, with a fund size of \$10,000, each employee receives \$1,000. But if the size of the fund is \$24,000, then the best worker gets \$6,000, and each of the others receives \$2,000.

The third strengthening of endowment monotonicity is also suggested by the examples from Table~\ref{table:golfpoker}. The prize structures from golf and poker are \emph{scale invariant}, that is, when the prize fund is increased $k$ times, the prize for each position also increases $k$ times. We show that the only two fair, consistent, and scale invariant rules for the distribution of prizes are the Equal Division rule and the Winner-Takes-All rule (Corollary~\ref{cor:EADD}). As a result, we see that the prize structures from poker and golf are not consistent.

Since consistency can be an excessive requirement, we further formulate two relaxations of consistency.\footnote{Our weakening of consistency was inspired by local stability introduced by \citet{Young88}.} Imagine that a tournament is held among participants of different skill levels. The organizer could distribute the prize fund according to the general ranking of all participants; or the organizer could divide the prize fund into two parts and distribute the prizes separately among the participants of high and low levels. If both methods lead to the same distribution of prizes, then the rule satisfies \emph{local consistency}. If the prize distribution is at least the same for high level participants, then the rule satisfies \emph{top consistency}. Such rules are particularly desirable in competitions where losers take their prizes and leave the competition one by one (as in poker) or in groups (as in golf). For example, Table~\ref{table:golfpoker} shows that the top 10 poker tournament participants receive a total of \$5.605 million of a total prize endowment of \$11.18 million. Local and top consistency require that if you apply the rule separately to the top 10 participants and a prize endowment of \$5.605 million, then they will receive the same prizes.

Our second main result describes the family of \emph{fair and locally consistent} rules that satisfy winner strict endowment monotonicity. We call them \emph{single-parametric rules}. Each such rule is generated by some continuous and non-decreasing function $f$ as follows. The winner of the competition receives some prize, $x$, the runner-up receives a prize $f(x)$, the third position receives a prize $f(f(x))$, and so on. The value of $x$ is unambiguously determined from the condition that the organizer distributes the whole prize endowment (Theorem~\ref{theorem:localcons}, Corollary~\ref{cor:lcons}). Further, we show that the prize structure is fair, locally consistent and scale invariant, if and only if the prizes form a \emph{geometric} sequence with representation $\lambda$, that is, the prize for position $r+1$ is the fraction $\lambda$ of the prize for position $r$ (Corollary~\ref{cor:geo}). For example, the prizes in the final stage of the poker tournament from Table~\ref{table:golfpoker} form a geometric sequence with $\lambda = 0.713$ and, therefore, are locally consistent.\footnote{The IBU World Cup Biathlon is another example, where the prizes are almost geometrical \citep{KondratevIanovskiNesterov19}.} The prizes in golf tournaments, however, do not form a geometric sequence and, therefore, are not even locally consistent.

Our third main result describes the family of \emph{fair and top consistent} rules that satisfy winner strict endowment monotonicity. Each such \emph{parametric rule} is generated by some sequence of continuous and non-decreasing functions $f_2, f_3, \ldots$, as follows. The winner of the competition receives some prize, $x$, the runner-up receives a prize $f_2(x)$, the third position receives a prize $f_3(x)$, and so on. The value of $x$ is unambiguously determined from the condition that the organizer distributes the whole prize endowment (Theorem~\ref{theorem:topcons}, Corollary~\ref{cor:topcons}). Finally, we show that the prize structure is fair, top consistent and scale invariant, if and only if there exists some sequence of $\lambda_1, \lambda_2, \ldots$, such that the prize for position $r$ is the fraction $\lambda_r/\lambda_1$ of the prize for the winner (Corollary~\ref{cor:topprop}). We call them \emph{proportional rules}. For example, the prizes in the golf tournaments from Table~\ref{table:golfpoker} are distributed according to a proportional rule with $\lambda_1=18.0, \lambda_2=10.9, \lambda_3=6.9$, and so on.

Our results are presented in Tables~\ref{table:properties} and~\ref{table:properties2}. Note that no prize allocation rule satisfies all properties considered. However, the geometric rules with representation $\lambda$ such that $0<\lambda<1$ satisfy all properties except for consistency.

\begin{table}[ht!]
\caption{Properties of consistent rules}\label{table:properties}
{\small
\begin{center}\begin{tabular}{|c|c|c|c|c|c|}
\hline
&  & Winner- & & Winner- & \\
&  & Takes- & Equal & Takes- & WTA \\
& Interval & All & Division & Surplus & and \\
& Rules & (WTA) & (ED) & (WTS) & ED \\
\hline
&  Th.~\ref{theorem:cons} & Cor.~\ref{cor:sop} & Cor.~\ref{cor:ED} & Cor.~\ref{cor:WTS} & Cor.~\ref{cor:EADD} \\
\hline
& $(a_k, b_k),$ & one & one & $a\in[0,\infty]$ & two\\
& $k=1,2,\ldots$ & rule & rule & & rules \\
\hline
Anonymity & + & + & + & + & + \\
\hline
Order Preservation & +* & + & +* & +* & +* \\
\hline
Winner-Loser Strict & only & +* & $-$ & only & only \\
Order Preservation & WTA & & & WTA & WTA \\
\hline
Strict Order Preservation & $-$ & $-$ & $-$ & $-$ & $-$ \\
\hline
Endowment Monotonicity & +* & +* & + & + & + \\
\hline
Winner Strict & only & + & + & +* & + \\
Endowment Monotonicity & WTS &  &  &  & \\
\hline
Strict & only & $-$ & +* & only & only \\
Endowment Monotonicity & ED & & & ED & ED \\
\hline
Scale Invariance & only & + & + & only & +* \\
(Endowment Additivity) & WTA,ED & & & WTA,ED & \\
\hline
Top Consistency & + & + & + & + & + \\
\hline
Local Consistency & + & + & + & + & + \\
\hline
Consistency & +* & +* & +* & +* & +* \\
\hline
\end{tabular}\end{center}}
\footnotesize{\emph{Notes}:
\begin{itemize}
    \item + indicates that the property is satisfied by each rule in the family, $-$ indicates that the property is not satisfied by each rule in the family, and * indicates the axiomatic characterizations;
    \item a WTS rule with representation $a$ is an interval rule with $a_1=a$ and $b_1=\infty$;
    \item the WTA rule is an interval rule with $a_1=0$, $b_1=\infty$, and a WTS rule with $a=0$;
    \item the ED rule is an interval rule with $a_k=b_k=0$ for all $k$, and a WTS rule with $a=\infty$.
\end{itemize}}
\end{table}

\begin{table}[ht!]
\caption{Properties of top and locally consistent rules}\label{table:properties2}
{\small
\begin{center}\begin{tabular}{|c|c|c|c|c|}
\hline
& Single- & & &\\
& Parametric & Geometric & Parametric & Proportional \\
& Rules & Rules & Rules & Rules \\
\hline
& Th.~\ref{theorem:localcons} & Cor.~\ref{cor:geo} & Th.~\ref{theorem:topcons} & Cor.~\ref{cor:topprop} \\
& Cor.~\ref{cor:lcons} & & Cor.~\ref{cor:topcons} & \\
\hline
& function $f$: & $\lambda\in [0,1]$ & functions $f_k$: & $\lambda_k\in[0,\infty)$: \\
& $0\leq f(x)\leq x$ & & $0\leq f_{k+1}(x)\leq f_k(x)\leq x$ & $\lambda_{k+1}\leq\lambda_k$ \\
& & & $k=1,2,\ldots$ & $k=1,2,\ldots$ \\
\hline
Anonymity & +* & +* & +* & +* \\
\hline
Order Preservation & +* & +* & +* & +* \\
\hline
Winner-Loser Strict & $f(x)<x$ & all but & $f_2(x)<x$ & $\lambda_2<\lambda_1$ \\
Order Preservation & for $x>0$ & ED & for $x>0$ & \\
\hline
Strict  & $0<f(x)<x$ & all but & $f_{k+1}(x)<f_k(x)$ & $\lambda_{k+1}<\lambda_k$ \\
Order Preservation & for $x>0$ & WTA,ED & for $x>0$, for each $k$ & for each $k$ \\
\hline
Endowment Monotonicity & + & + & + & + \\
\hline
Winner Strict & +* & + & +* & + \\
Endowment Monotonicity & & & & \\
\hline
Strict & $f$ is & all but & $f_k$ is increasing & $\lambda_k>0$ \\
Endowment Monotonicity & increasing & WTA & for each $k$ & for each $k$ \\
\hline
Scale Invariance & only & +* & only & +* \\
(Endowment Additivity) & Geometric & & Proportional & \\
\hline
Top Consistency & + & + & +* & +* \\
\hline
Local Consistency & +* & +* & only & only \\
& & & Single-Parametric & Geometric \\
\hline
Consistency & only & only & only & only \\
 & WTS & WTA,ED & WTS & WTA,ED \\
\hline
\end{tabular}\end{center}}
\footnotesize{\emph{Notes}:
\begin{itemize}
    \item + indicates that the property is satisfied by each rule in the family, and * indicates the axiomatic characterizations;
    \item the Winner-Takes-All rule (WTA) is a single-parametric rule with $f(x)=0$ for each~$x$, a geometric rule with $\lambda=0$, a parametric rule with $f_1(x)=x$ and $f_k(x)=0$ for each $k\geq 2$, and a proportional rule with $\lambda_1=1$ and $\lambda_k=0$ for each $k\geq 2$;
    \item the Equal Division rule (ED) is a single-parametric rule with $f(x)=x$ for each~$x$, a geometric rule with $\lambda=1$, a parametric rule with $f_k(x)=x$ for each $k$, and a proportional rule with $\lambda_k=1$ for each $k$;
    \item a Winner-Takes-Surplus rule (WTS) with representation $a$ is a single-parametric rule with $f(x)=\min\{a,x\}$ for each~$x$, and a parametric rule with $f_1(x)=x$ and $f_k(x)=\min\{a,x\}$ for each $k\geq 2$;
    \item a geometric rule with representation $\lambda$ is a single-parametric rule with $f(x)=\lambda x$ for each~$x$, a parametric rule with $f_k(x)=\lambda^{k-1}x$ for each $k$, and a proportional rule with $\lambda_k=\lambda^{k-1}$ for each $k$;
    \item a single-parametric rule with representation $f$ is a parametric rule with $f_k=f^{(k-1)}$ for each $k$;
    \item a proportional rule with representation $\lambda_k$ is a parametric rule with $f_k(x)=\lambda_kx/\lambda_1$ for each~$k$.
\end{itemize}}
\end{table}

\subsection{Literature overview}\label{sec:overview}

In this brief literature overview, we highlight other papers related to our key topics: fairness, consistency, rank-order tournaments, prize allocation, and axiomatic characterization.

Francis \citet{Galton1902} was perhaps the first to write on how the total prize money should be divided into prizes for each ranking position in a rank-order tournament. Using a probabilistic approach with order statistics, Galton concluded that when only two prizes are given, the first prize should approximately be three times the value of the second. Because of poor economic motivation, this approach has not received noticeable attention in the literature.

The seminal paper of \citet{LazearRosen81} analyzed rank-order tournaments from an economic perspective. In their model, a firm assigns a certain prize to each ranking position. Then each worker chooses a level of effort which leads to an output. The relative ranking of the outputs determines the prizes for workers. Lazear and Rosen found the optimal prize structure that maximizes the worker's utility in equilibrium. The subsequent literature proposed similar models and studied optimal prize structures that maximize different goals, such as the total output \citep{GlazerHassin88}, the revenue to the auctioneer \citep{BarutKovenock98}, the total effort of competitors \citep{MoldovanuSela01}, the highest effort among all competitors \citep{MoldovanuSela06}, the weighted total effort of the top $k$ competitors \citep{ArchakSundararajan09}, the participation of competitors \citep{AzmatMoller09}, and the participation of talented competitors \citep{AzmatMoller18}. 

In the literature on rank-order tournaments, anonymity and order preservation are standard fairness properties for prize allocation rules. In the model of \citet{LazearRosen81}, a firm assigns a certain prize to each position regardless of the identity of the worker who occupies the position. \citet{OKeeffeViscusiZeckhauser84} studied anonymous prize allocation, which they call `fair', and compared it with non-anonymous prize allocation, which they call `unfair'. \citet{GlazerHassin88}, \citet{BarutKovenock98}, \citet{MoldovanuSela01}, \citet{MoldovanuSela06}, and \citet{AzmatMoller09} considered the order preservation of prizes as a model assumption. \citet{ArchakSundararajan09} call order preserving prizes `fair'. \citet{OlszewskiSiegel20} found that total effort maximization, concave prize valuations, and convex effort costs call for the strict order preservation of prizes. 

The Winner-Takes-Surplus (WTS) and geometric prize sequences that we axiomatically characterize appeared in the literature for other reasons. \citet{MoldovanuSelaShi07} introduced contests for status and found that when the total prize money is high enough, the WTS prize sequence maximizes the total effort of competitors. \citet{GershkovLiSchweinzer09} showed that under some assumptions the WTS prize sequence is an efficient redistribution of output among partners within teams. \citet{NewmanTafkov14} provided an experiment and showed that relative performance information has a negative effect on performance of competitors if the WTS prize sequence is used. \citet{Xiao16} studied geometric prize sequences because under some assumptions it guarantees the existence and uniqueness of a Nash equilibrium, which is hard to prove for general prize sequences.

The consistency principle for allocation rules was motivated by \citet{BalinskiYoung82} as ``every part of a fair division should be fair''. Later, \citet{Thomson2012} argued that this should be restated as ``every part of every socially desirable allocation should be socially desirable''. Consistency has been applied in several models for allocation problems with infinitely divisible goods. \citet{Moulin1985} characterized equal cost allocation rules using consistency under the name `separability'. Inspired by the Talmud, consistency was applied to bankruptcy problems \citep{AumannMaschler1985}, taxation problems \citep{Young1987}, and rationing problems \citep{Moulin2000}. Under the name `stability', \citet{Lensberg1987} studied the implications of consistency in the context of bargaining problems, in particular for the Nash solution \citep{Lensberg1988}. \citet{Thomson1988} used consistency to characterize the Walrasian correspondence for exchange economies. The core and the Shapley value for cooperative games were characterized in terms of consistency by \citet{Peleg1986} and \citet{HartMasColell1989}, respectively. \citet{Thomson1994} showed that the central uniform rule for allocation problems with single-peaked preferences admits a characterization based on consistency. Consistent solutions in atomless economies were analyzed by \citet{ThomsonZhou1993}. Recently, \citet{Hougaard2017} studied consistency in the context of revenue sharing for hierarchical ventures. \citet{Osorio2017} employed consistency for operational problems (such as river sharing problems, sequential allocation and rationing problems) and obtained a geometric structure for resource allocation.

We are aware of two applications of the consistency axiom in the context of competitions. First, for competitions in which participants put in effort to increase their probability of winning a single prize, \citet{Skaperdas96} characterized the additive contest success functions, which provide each player’s probability of winning as a function of all players’ efforts. Though this model is mathematically equivalent to prize allocation based on cardinal performance of competitors, we are not aware of real examples of competitions using such prize allocation schemes. Second, \citet{FloresSzwagrzakTreibich20} characterized measures of individual productivity when team membership and team production are observable but individual contributions to team production are not. The authors illustrated their approach with the data from the National Basketball League.

The literature on prize allocation in rank-order competitions has not taken an axiomatic approach and the fair allocation literature has not focused on competitions. We connect these two fields by applying the axiomatic approach to prize allocation in rank-order competitions. Surprisingly, the literature on this topic is scarce. A rare example is \citet{PetroczyCsato21revenue}, who proposed a parametric rule based on pairwise comparisons and illustrated their approach with the data from the FIA Formula One racing. However, the authors did not provide axiomatic characterizations. To our knowledge, our paper is the first that develops and motivates prize allocation rules for rank-order competitions directly from axioms.\footnote{Recently, \citet{BergantinosMorenoTernero20JME,BergantinosMorenoTernero20MS,BergantinosMorenoTernero21jebo,BergantinosMorenoTernero21ejor} initiated axiomatic studies on broadcasting revenue sharing, which is another essential part of money sharing in competitions (e.g., in soccer leagues).}

This paper is structured as follows. Section~\ref{sec:cons} introduces the model and characterizes fair and consistent prize allocation rules. Section~\ref{sec:localcons} characterizes fair and locally consistent rules, and fair and top consistent rules. Section~\ref{sec:conrem} concludes. The proofs are contained in the Appendix.

\section{Consistent prize allocation rules}\label{sec:cons}

\subsection{Model}

Let $U$ be a countable set of at least three potential \emph{competitors}. On each given occasion, a finite subset $N\subseteq U$ is in a competition. This competition results in a \emph{ranking}, a bijection $\mathcal{R}:N\to\{1,\ldots,|N|\}$ assigning to each competitor a \emph{position}. Here, competitor $i\in N$ has a higher position in the ranking than competitor $j\in N$ if $\mathcal{R}(i)<\mathcal{R}(j)$. The \emph{prize endowment} $E\in\mathbb{R}_+$ is the amount of prize money to be allocated among the competitors. Thus, a \emph{competition} is a triple $(N,\mathcal{R},E)$.

Assuming that more money is better for each competitor, the preferences of the competitors over the feasible prize allocations are in conflict. In order to select a reasonable compromise for each competition, we study prize allocation rules $\varphi$ which assign to each competition $(N,\mathcal{R},E)$ an allocation of the prize endowment among the competitors, i.e.\ $\varphi(N,\mathcal{R},E)\in\mathbb{R}_+^N$ is such that 
\begin{equation*}
    \sum_{i\in N}\varphi_i(N,\mathcal{R},E)=E.
\end{equation*}
Throughout this paper, $\varphi$ denotes a generic prize allocation rule.

Reflecting the opposing principles of egalitarianism and elitism, two extreme and basic prize allocation rules are the \emph{Equal Division rule} and the \emph{Winner-Takes-All rule}. The Equal Division rule divides the prize money equally among the competitors. The Winner-Takes-All rule allocates all the prize money to the competitor with the highest position.

\begin{property}{Equal Division rule (ED)}\ \\
The equal division rule ${\rm ED}$ assigns to each competition $(N,\mathcal{R},E)$ the prize allocation
\begin{equation*}
    {\rm ED}_i(N,\mathcal{R},E)=\frac{E}{|N|} \quad \text{for each $i\in N$.}
\end{equation*}
\end{property}

\begin{property}{Winner-Takes-All rule (WTA)}\ \\
The winner-takes-all rule ${\rm WTA}$ assigns to each competition $(N,\mathcal{R},E)$ the prize allocation
\begin{equation*}
    {\rm WTA}_i(N,\mathcal{R},E)=
    \begin{cases}
        E & \text{if $\mathcal{R}(i)=1$;} \\
        0 & \text{otherwise.}
    \end{cases}
\end{equation*}
\end{property}

The ED rule is fair from an egalitarian perspective since it treats each competitor equally; the WTA rule is fair from an elitist perspective since it rewards the winner for achieving the highest position in the ranking.
\\[2ex]
\indent We take an axiomatic approach to study the fundamental differences between prize allocation rules. This means that we formulate some desirable properties of rules and analyze their implications. An elementary property imposing a form of equal treatment of competitors is \emph{anonymity}, which requires that the prize allocation does not depend on the identities of the competitors, but only on the number of competitors, their ranking, and the prize endowment.

\begin{property}{Axiom: anonymity}\ \\
For each pair of competitions $(N,\mathcal{R},E)$ and $(N',\mathcal{R}',E)$ with equal numbers of competitors $|N|=|N'|$, and each pair of competitors $i\in N$ and $j\in N'$ with equal positions $\mathcal{R}(i)=\mathcal{R}'(j)$, we have
\begin{equation*}
    \varphi_i(N,\mathcal{R},E)=\varphi_j(N',\mathcal{R}',E).
\end{equation*}
\end{property}

The fairness property imposing the prize allocation to reflect the ranking is \emph{order preservation}, which requires that the prize of a competitor is not lower than the prize of a competitor with a lower position in the ranking.

\begin{property}{Axiom: order preservation}\ \\
For each competition $(N,\mathcal{R},E)$, if competitor $i\in N$ has a higher position in the ranking $\mathcal{R}$ than competitor $j\in N$, we have
\begin{equation*}
    \varphi_i(N,\mathcal{R},E)\geq\varphi_j(N,\mathcal{R},E).
\end{equation*}
\end{property}

The solidarity property \emph{endowment monotonicity} requires that no competitor is worse off when the prize endowment increases.

\begin{property}{Axiom: endowment monotonicity}\ \\
For each pair of competitions $(N,\mathcal{R},E)$ and $(N,\mathcal{R},E')$ with $E<E'$ and each competitor $i\in N$, we have
\begin{equation*}
    \varphi_i(N,\mathcal{R},E)\leq\varphi_i(N,\mathcal{R},E').
\end{equation*}
\end{property}

Endowment monotonicity is a stronger property than \emph{endowment continuity}, which requires that small changes in the prize endowment have a small impact on the assigned prize allocation.

\begin{property}{Axiom: endowment continuity}\ \\
For each pair of competitions $(N,\mathcal{R},E)$ and $(N,\mathcal{R},E')$, we have $\varphi(N,\mathcal{R},E)\rightarrow \varphi(N,\mathcal{R},E')$ if $E\rightarrow E'$.
\end{property}

\begin{lemma}\label{lemma:econt}\ \\
If a prize allocation rule satisfies endowment monotonicity, then it satisfies endowment continuity.
\end{lemma}

\noindent All proofs are provided in the Appendix.

\subsection{Joint characterization}

Prize allocation may depend on the number of competitors. A criterion for evaluating whether a prize allocation rule prescribes coherent allocations for competitions with different numbers of competitors is \emph{consistency}. Consider an arbitrary competition $(N,\mathcal{R},E)$ and suppose that some competitors $S\subseteq N$ redistribute their accumulated prizes. This redistribution is based on their \emph{subranking}, the bijection $\mathcal{R}_S:S\to\{1,\ldots,|S|\}$ such that for each pair of competitors $i\in S$ and $j\in S$, we have $\mathcal{R}_S(i)<\mathcal{R}_S(j)$ if and only if $\mathcal{R}(i)<\mathcal{R}(j)$. A prize allocation rule is consistent if it allocates to each competitor $i\in S$ in the corresponding reduced competition the same prize as in the original competition.

\begin{property}{Axiom: consistency}\ \\
For each competition $(N,\mathcal{R},E)$, each nonempty subset of competitors $S\subseteq N$, and each competitor $i\in S$, we have\footnote{In fact, all results in this section hold when consistency is weakened to \emph{bilateral consistency}, requiring consistency only for reduced competitions with two competitors.}
\begin{equation*}
    \varphi_i(N,\mathcal{R},E)=\varphi_i\left(S,\mathcal{R}_S,\sum_{j\in S}\varphi_j(N,\mathcal{R},E)\right).
\end{equation*}
\end{property}

The Equal Division rule and the Winner-Takes-All rule both satisfy anonymity, order preservation, endowment monotonicity, and consistency, but these rules are not the only ones. The following example provides another rule satisfying these properties.

\begin{example}\label{ex:step}\ \\
{\rm Let a prize allocation rule be defined by allocating the prize endowment in the following way. Up to the first dollar is allocated to the competitor with the highest position, the surplus up to the second dollar is allocated to the competitor with the second highest position, and so on until each competitor has been allocated one dollar. If there is still money left, the first additional dollar is allocated to the competitor with the highest position, the second additional dollar to the competitor with the second highest position, and so on. This procedure continues until the full prize endowment is allocated among the competitors. Table~\ref{table:example1} illustrates how a prize endowment from one to six dollars is allocated among two, three, and four competitors. This rule also satisfies anonymity, order preservation, endowment monotonicity, and consistency.
\closeex}
\end{example}

\begin{table}[ht!]
\caption{Six dollars allocated among two, three, and four competitors in Example~\ref{ex:step}}\label{table:example1}
{\small
\begin{center}
\begin{tabular}{cc|c}
\toprule
$\varphi_1$&	$\varphi_2$& $E$\\
\hline
1&	0&	1\\
1&	1&	2\\
2&	1&	3\\
2&	2&	4\\
3&	2&	5\\
3&	3&	6\\
\bottomrule
\end{tabular}
\quad
\begin{tabular}{ccc|c}
\toprule
$\varphi_1$&	$\varphi_2$&	$\varphi_3$&    $E$\\
\hline
1&	0&	0&	1\\
1&	1&	0&	2\\
1&	1&	1&	3\\
2&	1&	1&	4\\
2&	2&	1&	5\\
2&	2&	2&	6\\
\bottomrule
\end{tabular}
\quad
\begin{tabular}{cccc|c}
\toprule
$\varphi_1$&	$\varphi_2$&	$\varphi_3$&	$\varphi_4$& $E$\\
\hline
1&	0&	0&	0&	1\\
1&	1&	0&	0&	2\\
1&	1&	1&	0&	3\\
1&	1&	1&	1&	4\\
2&	1&	1&	1&	5\\
2&	2&	1&	1&	6\\
\bottomrule
\end{tabular}
\end{center}}
\end{table}

It turns out that anonymity is implied by order preservation, endowment monotonicity, and consistency. This is captured by the following lemma.

\begin{lemma}\label{lemma:anonymity}\ \\
If a prize allocation rule satisfies order preservation, endowment monotonicity, and consistency, then it satisfies anonymity.
\end{lemma}

To have a full understanding of the joint implication of order preservation, endowment monotonicity, and consistency, we only need to focus on the structure of the corresponding prize allocation rules for competitions with two competitors. The following lemma shows that each such rule has at most one consistent extension.\footnote{A similar result for consistent extensions in the context of claims problems was obtained by \citet{AumannMaschler1985}.}

\begin{lemma}\label{lemma:cons}\ \\
If two prize allocation rules satisfying endowment monotonicity and consistency coincide for each competition with two competitors, then the two rules coincide for each competition with an arbitrary number of competitors.
\end{lemma}

How do prize allocation rules satisfying order preservation, endowment monotonicity, and consistency look for competitions with two competitors? To this end, it helps to graphically illustrate possible \emph{prize allocation paths}, i.e.\ draw the prize allocations assigned to competitions with two competitors when the prize endowment increases from zero.

\setcounter{example}{0}
\begin{example}[continued]\ \\
{\rm Let $(N,\mathcal{R},E)$ with $|N|=2$ be a competition with two competitors. Denote $N=\{1,2\}$ such that $\mathcal{R}(1)=1$ and $\mathcal{R}(2)=2$. Let the prize endowment $E$ gradually increase from zero. Then the prize allocation paths of the ED rule, the WTA rule, and the prize allocation rule $\varphi$ described above are illustrated as follows.
\begin{center}\begin{tikzpicture}
\draw[->] (0,0) -- (5,0) node[anchor=north] {$\varphi_1$};
\draw	(0,0) node[anchor=north] {0}
        (1,0) node[anchor=north] {1}
        (2,0) node[anchor=north] {2}
        (3,0) node[anchor=north] {3}
		(4,0) node[anchor=north] {4};
\draw[->] (0,0) -- (0,5) node[anchor=east] {$\varphi_2$};
\draw	(0,1) node[anchor=east] {1}
		(0,2) node[anchor=east] {2}
		(0,3) node[anchor=east] {3}
        (0,4) node[anchor=east] {4};
\draw[dashed] (1,0) -- (0,1);
\draw[dashed] (2,0) -- (0,2);
\draw[dashed] (3,0) -- (0,3);
\draw[dashed] (4,0) -- (0,4);
\draw[dashed] (5,0) -- (0,5);
\draw[dashed] (5,1) -- (1,5);
\draw[dashed] (5,2) -- (2,5);
\draw[dashed] (5,3) -- (3,5);
\draw[dashed] (5,4) -- (4,5);
\draw[thick] (0,0) -- (5,5);
\draw (3.2,4.5) node {${\rm ED}(N,\mathcal{R},E)$};
\draw[thick] (0,0) -- (5,0);
\draw (3.7,0.5) node {${\rm WTA}(N,\mathcal{R},E)$};
\draw[very thick] (0,0) -- (1,0);
\draw[very thick] (1,0) -- (1,1);
\draw[very thick] (1,1) -- (2,1);
\draw[very thick] (2,1) -- (2,2);
\draw[very thick] (2,2) -- (3,2);
\draw[very thick] (3,2) -- (3,3);
\draw[very thick] (3,3) -- (4,3);
\draw[very thick] (4,3) -- (4,4);
\draw[very thick] (4,4) -- (5,4);
\draw[very thick] (5,4) -- (5,5);
\draw (4.1,2.5) node {$\varphi(N,\mathcal{R},E)$};
\end{tikzpicture}
\end{center}
The horizontal axis depicts the prize allocated to competitor one and the vertical axis depicts the prize allocated to competitor two. The dotted lines indicate different levels of the prize endowment.
\closeex}
\end{example}

Each prize allocation rule satisfying order preservation, endowment monotonicity, and consistency is somehow a combination of the ED rule, the WTA rule, and the type of rule in Example~\ref{ex:step}. Each such prize allocation rule can be described in the following way. Let $N\subseteq U$ be a finite set of competitors and consider a corresponding ranking and prize endowment. Then there exist disjoint intervals $(a_k,b_k)$ with $a_k\in\mathbb{R}_+$ and $b_k\in\mathbb{R}_+\cup\{+\infty\}$ for each $k$ such that the following holds. Each competitor is allocated a prize of $a_k$ when the prize endowment equals $|N|a_k$. If the endowment is higher, first the prize allocated to the competitor with the highest position is increased to $b_k$, then the prize allocated to the competitor with the second highest position is increased to $b_k$, and so on. This means that each competitor is allocated a prize of $b_k$ when the prize endowment equals $|N|b_k$. If the average prize endowment does not belong to one of these intervals, it is equally divided among the competitors. We call such prize allocation rules \emph{interval rules}.

\begin{property}{Interval rules}\ \\
There exist disjoint intervals $(a_1,b_1),(a_2,b_2),\ldots$ with $a_1,a_2,\ldots\in\mathbb{R}_+$ and $b_1,b_2,\ldots\in\mathbb{R}_+\cup\{+\infty\}$ such that for each competition $(N,\mathcal{R},E)$ and each competitor $i\in N$, we have
\begin{equation*}
    \varphi_i(N,\mathcal{R},E)=
    \begin{cases}
        a_k & \text{if $|N|a_k\leq E\leq(|N|-\mathcal{R}(i)+1)a_k+(\mathcal{R}(i)-1)b_k$};
    \\  x & \text{if $(|N|-\mathcal{R}(i)+1)a_k+(\mathcal{R}(i)-1)b_k\leq E\leq(|N|-\mathcal{R}(i))a_k+\mathcal{R}(i)b_k$};
    \\  b_k & \text{if $(|N|-\mathcal{R}(i))a_k+\mathcal{R}(i)b_k\leq E\leq|N|b_k$};
    \\  \frac{E}{|N|} & \text{otherwise},
    \end{cases}
\end{equation*}
where $x=E-(|N|-\mathcal{R}(i))a_k-(\mathcal{R}(i)-1)b_k$.
\end{property}

Note that the ED rule is an interval rule with $a_k=b_k=0$ for each $k$. The WTA rule is an interval rule with $a_1=0$ and $b_1=+\infty$. The prize allocation rule from Example~\ref{ex:step} is an interval rule with $a_k=k-1$ and $b_k=k$ for each $k$.

\begin{example}\ \\
{\rm Let $(N,\mathcal{R},E)$ with $|N|=2$ be a competition with two competitors. Denote $N=\{1,2\}$ such that $\mathcal{R}(1)=1$ and $\mathcal{R}(2)=2$. The prize allocation path of the interval rule $\varphi$ with $a_1=1$, $b_1=a_2=2\frac{1}{2}$, $b_2=3$, $a_3=3\frac{1}{2}$, and $b_3=+\infty$ is illustrated as follows.
\begin{center}\begin{tikzpicture}
\draw[->] (0,0) -- (5,0) node[anchor=north] {$\varphi_1$};
\draw	(0,0) node[anchor=north] {0}
        (1,0) node[anchor=north] {1}
        (2,0) node[anchor=north] {2}
        (3,0) node[anchor=north] {3}
		(4,0) node[anchor=north] {4};
\draw[->] (0,0) -- (0,5) node[anchor=east] {$\varphi_2$};
\draw	(0,1) node[anchor=east] {1}
		(0,2) node[anchor=east] {2}
		(0,3) node[anchor=east] {3}
        (0,4) node[anchor=east] {4};
\draw[dashed] (1,0) -- (0.5,0.5);
\draw[dashed] (2,0) -- (1,1);
\draw[dashed] (3,0) -- (1.5,1.5);
\draw[dashed] (4,0) -- (2,2);
\draw[dashed] (5,0) -- (2.5,2.5);
\draw[dashed] (5,1) -- (3,3);
\draw[dashed] (5,2) -- (3.5,3.5);
\draw[dashed] (5,3) -- (4,4);
\draw[dashed] (5,4) -- (4.5,4.5);
\draw[dashed] (0,0) -- (5,5);
\draw[very thick] (0,0) -- (1,1);
\draw[very thick] (1,1) -- (2.5,1);
\draw[very thick] (2.5,1) -- (2.5,2.5);
\draw[very thick] (2.5,2.5) -- (3,2.5);
\draw[very thick] (3,2.5) -- (3,3);
\draw[very thick] (3,3) -- (3.5,3.5);
\draw[very thick] (3.5,3.5) -- (5,3.5);
\draw (2,4) node {$\varphi(N,\mathcal{R},E)$};
\draw[dashed] (0,1) -- (1,1);
\draw[dashed] (0,2.5) -- (2.5,2.5);
\draw[dashed] (0,3) -- (3,3);
\draw[dashed] (0,3.5) -- (3.5,3.5);
\end{tikzpicture}\end{center}
\closeex}
\end{example}


\begin{theorem}\label{theorem:cons}\ \\
A prize allocation rule satisfies order preservation, endowment monotonicity, and consistency if and only if it is an interval rule.
\end{theorem}

\bigskip

\subsection{Strengthening the fairness properties}

Theorem~\ref{theorem:cons} shows that many rules satisfy order preservation, endowment monotonicity, and consistency. This means that there is room for imposing additional requirements. In particular, it may be possible to strengthen one of the fairness properties. Unfortunately, order preservation cannot be strengthened to \emph{strict order preservation}, which requires that the prize of a competitor is higher than the prize of a competitor with a lower position. However, it can be strengthened to the weaker property \emph{winner-loser strict order preservation}, which requires in addition to order preservation that the prize of the competitor with the highest position is higher than the prize of the competitor with the lowest position. The Winner-Takes-All rule is the only interval rule satisfying winner-loser strict order preservation.

\begin{property}{Axiom: strict order preservation}\ \\
For each competition $(N,\mathcal{R},E)$ with $E>0$, if competitor $i\in N$ has a higher position in the ranking $\mathcal{R}$ than competitor $j\in N$, we have
\begin{equation*}
    \varphi_i(N,\mathcal{R},E)>\varphi_j(N,\mathcal{R},E).
\end{equation*}
\end{property}

\begin{property}{Axiom: winner-loser strict order preservation}\ \\
For each competition $(N,\mathcal{R},E)$ with $E>0$, if competitor $i\in N$ has a higher position in the ranking $\mathcal{R}$ than competitor $j\in N$, we have
\begin{equation*}
    \varphi_i(N,\mathcal{R},E)\geq\varphi_j(N,\mathcal{R},E)
\end{equation*}
and if $\mathcal{R}(i)=1$ and $\mathcal{R}(j)=|N|$, then
\begin{equation*}
    \varphi_i(N,\mathcal{R},E)>\varphi_j(N,\mathcal{R},E).
\end{equation*}
\end{property}

\begin{corollary}\label{cor:sop}\
\begin{enumerate}[(i)]
    \item No prize allocation rule satisfies strict order preservation, endowment monotonicity, and consistency.
    \item The unique prize allocation rule satisfying winner-loser strict order preservation, endowment monotonicity, and consistency is the Winner-Takes-All rule.
\end{enumerate}
\end{corollary}

Instead of strengthening order preservation, it is also possible to strengthen endowment monotonicity. For instance, \emph{strict endowment monotonicity} requires that each competitor is better off when the prize endowment increases, or the weaker property \emph{winner strict endowment monotonicity} requires in addition to endowment monotonicity that the winner is better off when the prize endowment increases.

\begin{property}{Axiom: strict endowment monotonicity}\ \\
For each pair of competitions $(N,\mathcal{R},E)$ and $(N,\mathcal{R},E')$ with $E<E'$ and each competitor $i\in N$, we have
\begin{equation*}
    \varphi_i(N,\mathcal{R},E)<\varphi_i(N,\mathcal{R},E').
\end{equation*}
\end{property}

\begin{property}{Axiom: winner strict endowment monotonicity}\ \\
For each pair of competitions $(N,\mathcal{R},E)$ and $(N,\mathcal{R},E')$ with $E<E'$ and each competitor $i\in N$, we have
\begin{equation*}
    \varphi_i(N,\mathcal{R},E)\leq\varphi_i(N,\mathcal{R},E')
\end{equation*}
and if $\mathcal{R}(i)=1$, then
\begin{equation*}
    \varphi_i(N,\mathcal{R},E)<\varphi_i(N,\mathcal{R},E').
\end{equation*}
\end{property}

The Equal Division rule is the only interval rule which satisfies strict endowment monotonicity. Moreover, an interval rule satisfies winner strict endowment monotonicity if and only if it is a \emph{Winner-Takes-Surplus rule}, i.e.\ it coincides with the ED rule up to some value of the prize endowment, and it allocates the surplus according to the WTA rule.

\begin{corollary}\label{cor:ED}\ \\
The unique prize allocation rule satisfying order preservation, strict endowment monotonicity, and consistency is the Equal Division rule.
\end{corollary}

\begin{property}{Winner-Takes-Surplus rules (WTS)}\ \\
There exists $a\in\mathbb{R}_+\cup\{+\infty\}$ such that for each competition $(N,\mathcal{R},E)$ and each competitor $i\in N$, we have
    \begin{equation*}
        \varphi_i(N,\mathcal{R},E)=
        \begin{cases}
            E-(|N|-1)a & \text{if $E\geq|N|a$ and $\mathcal{R}(i)=1$};
        \\  a & \text{if $E\geq|N|a$ and $\mathcal{R}(i)\neq1$};
        \\  \frac{E}{|N|} & \text{otherwise}.
    \end{cases}
    \end{equation*}
\end{property}

Note that the ED rule is a WTS rule with $a=+\infty$. The WTA rule is a WTS rule with $a=0$. Table~\ref{table:WTS} illustrates how a WTS rule $\varphi$ allocates a prize endowment from $a$ to $6a$ among two, three, and four competitors. The corresponding prize allocation path is illustrated as follows.
\begin{center}\begin{tikzpicture}
\draw[->] (0,0) -- (5,0) node[anchor=north] {$\varphi_1$};
\draw	(0,0) node[anchor=north] {0}
        (2,0) node[anchor=north] {$a$};
\draw[->] (0,0) -- (0,4) node[anchor=east] {$\varphi_2$};
\draw	(0,2) node[anchor=east] {$a$};
\draw[dashed] (1,0) -- (0.5,0.5);
\draw[dashed] (2,0) -- (1,1);
\draw[dashed] (3,0) -- (1.5,1.5);
\draw[dashed] (4,0) -- (2,2);
\draw[dashed] (5,0) -- (2.5,2.5);
\draw[dashed] (5,1) -- (3,3);
\draw[dashed] (5,2) -- (3.5,3.5);
\draw[dashed] (5,3) -- (4,4);
\draw[dashed] (0,0) -- (4,4);
\draw[very thick] (0,0) -- (2,2);
\draw[very thick] (2,2) -- (5,2);
\draw (1.5,3) node {$\varphi(N,\mathcal{R},E)$};
\draw[dashed] (2,0) -- (2,2);
\draw[dashed] (0,2) -- (2,2);
\end{tikzpicture}\end{center}

\begin{table}[ht!]
\caption{Winner-Takes-Surplus rule for two, three, and four competitors}\label{table:WTS}
{\small
\begin{center}
\begin{tabular}{cc|c}
\toprule
$\varphi_1$&	$\varphi_2$& $E$\\
\hline
$a/2$&	$a/2$&	$a$\\
$a$&	$a$&	$2a$\\
$2a$&	$a$&	$3a$\\
$3a$&	$a$&	$4a$\\
$4a$&	$a$&	$5a$\\
$5a$&	$a$&	$6a$\\
\bottomrule
\end{tabular}
\quad
\begin{tabular}{ccc|c}
\toprule
$\varphi_1$&	$\varphi_2$&	$\varphi_3$&    $E$\\
\hline
$a/3$&	$a/3$&	$a/3$&	$a$\\
$2a/3$&	$2a/3$&	$2a/3$&	$2a$\\
$a$&	$a$&	$a$&	$3a$\\
$2a$&	$a$&	$a$&	$4a$\\
$3a$&	$a$&	$a$&	$5a$\\
$4a$&	$a$&	$a$&	$6a$\\
\bottomrule
\end{tabular}
\quad
\begin{tabular}{cccc|c}
\toprule
$\varphi_1$&	$\varphi_2$&	$\varphi_3$&	$\varphi_4$& $E$\\
\hline
$a/4$&	$a/4$&	$a/4$&	$a/4$&	$a$\\
$a/2$&	$a/2$&	$a/2$&	$a/2$&	$2a$\\
$3a/4$&	$3a/4$&	$3a/4$&	$3a/4$&	$3a$\\
$a$&	$a$&	$a$&	$a$&	$4a$\\
$2a$&	$a$&	$a$&	$a$&	$5a$\\
$3a$&	$a$&	$a$&	$a$&	$6a$\\
\bottomrule
\end{tabular}
\end{center}}
\end{table}

\begin{corollary}\label{cor:WTS}\ \\
A prize allocation rule satisfies order preservation, winner strict endowment monotonicity, and consistency if and only if it is a Winner-Takes-Surplus rule.
\end{corollary}

Alternatively, endowment monotonicity can be strengthened to \emph{endowment additivity}. Suppose that the prize endowment turns out to be larger than expected. Then there are two ways of proceeding. Either the initial allocation is cancelled and the rule is applied to the competition with the new prize endowment, or the rule is applied to the competition with the increment as the prize endowment and the resulting allocation is added to the initial allocation. A prize allocation rule satisfies endowment additivity if both ways of proceeding lead to the same prize allocation.

\begin{property}{Axiom: endowment additivity}\ \\
For each pair of competitions $(N,\mathcal{R},E)$ and $(N,\mathcal{R},E')$ and each competitor $i\in N$, we have
\begin{equation*}
    \varphi_i(N,\mathcal{R},E+E')=\varphi_i(N,\mathcal{R},E)+\varphi_i(N,\mathcal{R},E').
\end{equation*}
\end{property}

Endowment additivity is equivalent to \emph{scale invariance}, which requires that each position in the ranking of a competition is assigned a fixed proportion of the prize endowment. Even when the prize endowment is not known in advance, the competitors know which share corresponds to each position. Among all interval rules, only the ED rule and the WTA rule satisfy endowment additivity (scale invariance).

\begin{property}{Axiom: scale invariance}\ \\
For each competition $(N,\mathcal{R},E)$ and each competitor $i\in N$, we have\footnote{The scale invariance axiom is also called \emph{homogeneity}.}
\begin{equation*}
    \varphi_i(N,\mathcal{R},E)=E\varphi_i(N,\mathcal{R},1).
\end{equation*}
\end{property}

\begin{lemma}\label{lemma:homo}\ \\
A prize allocation rule satisfies endowment additivity if and only if it satisfies scale invariance.
\end{lemma}

\begin{corollary}\label{cor:EADD}\ \\
The only two prize allocation rules satisfying order preservation, endowment additivity (scale invariance), and consistency are the Equal Division rule and the Winner-Takes-All rule.
\end{corollary}

\section{Weakening consistency}\label{sec:localcons}

\subsection{Locally consistent prize allocation rules}

We know that the Equal Division rule and the Winner-Takes-All rule are two members of a family of prize allocation rules satisfying anonymity, order preservation, endowment monotonicity, and consistency. Strengthening order preservation or endowment monotonicity leads to impossibilities, uniqueness, or restricted families. However, consistency may be considered a too strong requirement, since an arbitrary subranking may not reflect the results of the original competition well. Suppose for instance that two competitors of a competition with a large number of competitors reevaluate their allocated prizes on the basis of their subranking. Then the reduced competition tends to lose significant features of original competition, e.g.\ it does not take into account whether the two competitors originally ended up with the highest position and the second highest position, or with the highest position and the lowest position.

In fact, it may be desirable to weaken consistency to \emph{local consistency}, which requires only consistent allocations for reduced competitions where for each two competitors, each other competitor with an intermediate position is also involved. In other words, local consistency requires invariance under splitting up the full competition into leagues, where the competitors in the top segment of the ranking are put in a separate league, the competitors in the second segment of the ranking are put in a second league, and so on.

\begin{property}{Axiom: local consistency}\ \\
For each competition $(N,\mathcal{R},E)$, each subset of competitors $S\subseteq N$ with $|\mathcal{R}(i)-\mathcal{R}(j)|\leq|S|-1$ for all $i,j\in S$, and each competitor $i\in S$, we have\footnote{In fact, all results in this section hold when local consistency is weakened to \emph{bilateral local consistency}, requiring local consistency only for reduced competitions with two competitors.}
\begin{equation*}
    \varphi_i(N,\mathcal{R},E)=\varphi_i\left(S,\mathcal{R}_S,\sum_{j\in S}\varphi_j(N,\mathcal{R},E)\right).
\end{equation*}
\end{property}

In contrast to consistency, the following example shows that a prize allocation rule satisfying order preservation, endowment monotonicity, and local consistency does not necessarily satisfy anonymity.

\begin{example}\label{ex:anonymity}\ \\
{\rm Let $i\in U$ and $j\in U$ be two potential competitors. Let the prize allocation rule $\varphi$ be defined in the following way. If $i$ has the highest position and $j$ has the second highest position in the ranking, then $\varphi$ divides the prize endowment equally among competitors $i$ and $j$. Otherwise, $\varphi$ divides the prize endowment according to the WTA rule. Formally, $\varphi$ assigns to each competition $(N,\mathcal{R},E)$ the prize allocation such that\\ if $i,j\in N$, $\mathcal{R}(i)=1$, and $\mathcal{R}(j)=2$, then
\begin{equation*}
    \varphi_k(N,\mathcal{R},E)=
    \begin{cases}
        \tfrac{1}{2}E & \text{if $k\in\{i,j\}$};
    \\  0 & \text{otherwise},
    \end{cases}
\end{equation*}
and otherwise
\begin{equation*}
    \varphi_k(N,\mathcal{R},E)=
    \begin{cases}
        E & \text{if $\mathcal{R}(k)=1$};
    \\  0 & \text{otherwise}.
    \end{cases}
\end{equation*}
Then $\varphi$ satisfies order preservation, endowment monotonicity, and local consistency, but does not satisfy anonymity. By Lemma~\ref{lemma:anonymity}, $\varphi$ does not satisfy consistency.
\closeex}
\end{example}

Needless to say, all interval rules satisfy anonymity, order preservation, endowment monotonicity, and local consistency, but these rules are not the only ones. The following example provides another rule satisfying these properties.

\begin{example}\label{ex:arithmetic}\ \\
{\rm Let a prize allocation rule be defined by allocating the prize endowment in the following way. Up to the first dollar is allocated to the competitor with the highest position. The surplus is divided equally among the competitors with the highest position and the second highest position until they are allocated two and one dollar, respectively. Then the surplus is divided equally among the competitors with the three highest positions until they are allocated three, two, and one dollar, respectively. This procedure continues until the competitor with the lowest position is allocated one dollar. If there is still money left, this is equally divided among the competitors. Table~\ref{table:arithmetic} illustrates how a prize endowment from one to eight dollars is allocated among two, three, and four competitors. This rule also satisfies anonymity, order preservation, endowment monotonicity, and local consistency.

Let $(N,\mathcal{R},E)$ with $|N|=2$ be a competition with two competitors. Denote $N=\{1,2\}$ such that $\mathcal{R}(1)=1$ and $\mathcal{R}(2)=2$. Then the prize allocation paths of the ED rule, the WTA rule, and the prize allocation rule $\varphi$ described above are illustrated as follows.
\begin{center}\begin{tikzpicture}
\draw[->] (0,0) -- (5,0) node[anchor=north] {$\varphi_1$};
\draw	(0,0) node[anchor=north] {0}
        (1,0) node[anchor=north] {1}
        (2,0) node[anchor=north] {2}
        (3,0) node[anchor=north] {3}
		(4,0) node[anchor=north] {4};
\draw[->] (0,0) -- (0,5) node[anchor=east] {$\varphi_2$};
\draw	(0,1) node[anchor=east] {1}
		(0,2) node[anchor=east] {2}
		(0,3) node[anchor=east] {3}
        (0,4) node[anchor=east] {4};
\draw[dashed] (1,0) -- (0,1);
\draw[dashed] (2,0) -- (0,2);
\draw[dashed] (3,0) -- (0,3);
\draw[dashed] (4,0) -- (0,4);
\draw[dashed] (5,0) -- (0,5);
\draw[dashed] (5,1) -- (1,5);
\draw[dashed] (5,2) -- (2,5);
\draw[dashed] (5,3) -- (3,5);
\draw[dashed] (5,4) -- (4,5);
\draw[thick] (0,0) -- (5,5);
\draw (3.2,4.5) node {${\rm ED}(N,\mathcal{R},E)$};
\draw[thick] (0,0) -- (5,0);
\draw (3.8,0.5) node {${\rm WTA}(N,\mathcal{R},E)$};
\draw[very thick] (0,0) -- (1,0);
\draw[very thick] (1,0) -- (5,4);
\draw (4.1,2) node {$\varphi(N,\mathcal{R},E)$};
\end{tikzpicture}\end{center}
\closeex}
\end{example}

\begin{table}[ht!]
\caption{Eight dollars allocated among two, three, and four competitors in Example~\ref{ex:arithmetic}}\label{table:arithmetic}
{\small
\begin{center}
\begin{tabular}{cc|c}
\toprule
$\varphi_1$&	$\varphi_2$& $E$\\
\hline
1&	0&	1\\
$3/2$&	$1/2$&	2\\
2&	1&	3\\
$5/2$&	$3/2$&	4\\
3&	2&	5\\
$7/2$&	$5/2$&	6\\
4& 3& 7 \\
9/2& 7/2& 8\\
\bottomrule
\end{tabular}
\quad
\begin{tabular}{ccc|c}
\toprule
$\varphi_1$&	$\varphi_2$&	$\varphi_3$&    $E$\\
\hline
1&	0&	0&	1\\
$3/2$&	$1/2$&	0&	2\\
2&	1&	0&	3\\
$7/3$&	$4/3$&	$1/3$&	4\\
$8/3$&	$5/3$&	$2/3$&	5\\
3&	2&	1&	6\\
10/3& 7/3& 4/3& 7\\
11/3& 8/3& 5/3& 8\\
\bottomrule
\end{tabular}
\quad
\begin{tabular}{cccc|c}
\toprule
$\varphi_1$&	$\varphi_2$&	$\varphi_3$&	$\varphi_4$& $E$\\
\hline
1&	0&	0&	0&	1\\
$3/2$&	$1/2$&	0&	0&	2\\
2&	1&	0&	0&	3\\
$7/3$&	$4/3$&	$1/3$&	0&	4\\
$8/3$&	$5/3$&	$2/3$&	0&	5\\
3&	2&	1&	0&	6\\
13/4& 9/4& 5/4& 1/4& 7\\
7/2& 5/2& 3/2& 1/2& 8\\
\bottomrule
\end{tabular}
\end{center}}
\end{table}

Note that the arithmetic type of rule in Example~\ref{ex:arithmetic} satisfies winner strict endowment monotonicity, i.e.\ the competitor with the highest position is better off when the prize endowment increases. Each prize allocation rule satisfying anonymity, order preservation, winner strict endowment monotonicity, and local consistency admits a compact indirect description. For each such rule there exists a continuous and non-decreasing function $f:\mathbb{R}_+\to\mathbb{R}_+$ with $f(x)\leq x$ for each $x$ such that the assigned prize allocation for each competition can be described in the following way. The competitor with the highest position is allocated a prize of $x\in\mathbb{R}_+$. The competitor with the second highest position is allocated a prize of $f(x)$. The competitor with the third highest position is allocated a prize of $f(f(x))$, and so on. In total, the full prize endowment is allocated among the competitors. We call such prize allocation rules \emph{single-parametric rules}.

\begin{property}{Single-Parametric rules}\ \\
There exists a continuous and non-decreasing function $f:\mathbb{R}_+\to\mathbb{R}_+$ with $f(x)\leq x$ for each $x$ such that for each competition $(N,\mathcal{R},E)$ and each competitor $i\in N$, we have
\begin{equation*}
    \varphi_i(N,\mathcal{R},E)=f^{(\mathcal{R}(i)-1)}(x),
\end{equation*}
where we denote $f^{(0)}(x)=x$ and $f^{(k)}(x)=f(f^{(k-1)}(x))$ for each $k$, and $x\in\mathbb{R}_+$ is chosen such that $\sum_{k=1}^{|N|}f^{(k-1)}(x)=E$.
\end{property}

Note that the ED rule is a single-parametric rule with $f(x)=x$ for each $x$. The WTA rule is a single-parametric rule with $f(x)=0$ for each $x$. A Winner-Takes-Surplus rule (WTS) with representation $a$ is a single-parametric rule with $f(x)=\min\{a,x\}$ for each $x$. The arithmetic type of rule from Example~\ref{ex:arithmetic} is a single-parametric rule with $f(x)=\max\{0,x-1\}$ for each $x$.

\bigskip

\begin{theorem}\label{theorem:localcons}\ \\
A prize allocation rule satisfies anonymity, order preservation, winner strict endowment monotonicity, and local consistency if and only if it is a single-parametric rule.
\end{theorem}

\bigskip

Theorem~\ref{theorem:localcons} does not describe all prize allocation rules satisfying anonymity, order preservation, endowment monotonicity, and local consistency. If such rule does not satisfy winner strict endowment monotonicity, the prize of the competitor with the second highest position cannot be expressed as a function of the prize of the competitor with the highest position. Moreover, as the following example shows, such a rule for competitions with two competitors does not necessarily have a unique locally consistent extension to competitions with more competitors.

\begin{example}\label{ex:emon}\ \\
{\rm Let a prize allocation rule be defined by allocating the prize endowment in the following way. Up to the first dollar is allocated to the competitor with the highest position. Subsequently, one dollar is allocated to the competitor with the second highest position and then another dollar is allocated to the competitor with the highest position. Thereafter, one dollar is allocated to the competitor with the third highest position, then another dollar to the competitor with the second highest position, and then another dollar to the competitor with the highest position. This procedure continues until the competitor with the highest position is allocated an amount of dollars equal to the number of competitors. If there is still money left, first another dollar is allocated to the competitor with the lowest position, then another dollar to the competitor with the second lowest position, and so on. This continues until the full prize endowment is allocated among the competitors. Table~\ref{table:example5} illustrates how a prize endowment from one to eight dollars is allocated among two, three, and four competitors. This rule satisfies anonymity, order preservation, endowment monotonicity, and local consistency, but does not satisfy winner strict endowment monotonicity. Moreover, for competitions with two competitors, this rule coincides with the prize allocation rule from Example~\ref{ex:step}.
\closeex}
\end{example}

\begin{table}[ht!]
\caption{Eight dollars allocated among two, three, and four competitors in Example~\ref{ex:emon}}\label{table:example5}
{\small
\begin{center}
\begin{tabular}{cc|c}
\toprule
$\varphi_1$&	$\varphi_2$& $E$\\
\hline
1&	0&	1\\
1&	1&	2\\
2&	1&	3\\
2&	2&	4\\
3&	2&	5\\
3&	3&	6\\
4& 3& 7\\
4& 4& 8\\
\bottomrule
\end{tabular}
\quad
\begin{tabular}{ccc|c}
\toprule
$\varphi_1$&	$\varphi_2$&	$\varphi_3$&    $E$\\
\hline
1&	0&	0&	1\\
1&	1&	0&	2\\
2&	1&	0&	3\\
2&	1&	1&	4\\
2&	2&	1&	5\\
3&	2&	1&	6\\
3& 2& 2& 7\\
3& 3& 2& 8\\
\bottomrule
\end{tabular}
\quad
\begin{tabular}{cccc|c}
\toprule
$\varphi_1$&	$\varphi_2$&	$\varphi_3$&	$\varphi_4$& $E$\\
\hline
1&	0&	0&	0&	1\\
1&	1&	0&	0&	2\\
2&	1&	0&	0&	3\\
2&	1&	1&	0&	4\\
2&	2&	1&	0&	5\\
3&	2&	1&	0&	6\\
3& 2& 1& 1& 7\\
3& 2& 2& 1& 8\\
\bottomrule
\end{tabular}
\end{center}}
\end{table}

Subfamilies of single-parametric rules are obtained if order preservation is strengthened to strict order preservation or winner-loser strict order preservation, or winner strict endowment monotonicity is strengthened to strict endowment monotonicity. A single-parametric rule satisfies winner-loser strict order preservation if and only if it does not coincide with the ED rule for each positive prize endowment, and satisfies strict order preservation if and only if it does not coincide with the ED rule nor with the WTA rule for each positive prize endowment. Moreover, a single-parametric rule satisfies strict endowment monotonicity if and only if it does not allocate any additional prize endowment according to the WTA rule.

\begin{corollary}\label{cor:lcons}\ \\
Let $\varphi$ be a single-parametric rule with representation $f$. Then the following statements hold.
\begin{enumerate}[(i)]
    \item $\varphi$ satisfies winner-loser strict order preservation if and only if $f(x)<x$ for each $x>0$;
    \item $\varphi$ satisfies strict order preservation if and only if $0<f(x)<x$ for each $x>0$;
    \item $\varphi$ satisfies strict endowment monotonicity if and only if $f$ is increasing.
\end{enumerate}
\end{corollary}

If winner strict endowment monotonicity is strengthened to endowment additivity (scale invariance), then the interesting family of \emph{geometric rules} is obtained. For each geometric rule, there exists $\lambda\in[0,1]$ such that the assigned prize allocation for each competition can be described in the following way. The competitor with the second highest position is allocated a prize of $\lambda$ times the prize of the competitor with the highest position. The competitor with the third highest position is allocated a prize of $\lambda$ times the prize of the competitor with the second highest position, i.e.\ $\lambda^2$ times the prize of the competitor with the highest position. The competitor with the fourth highest position is allocated a prize of $\lambda^3$ times the prize of the competitor with the highest position, and so on. In total, the full prize endowment is allocated.

\begin{property}{Geometric rules}\ \\
There exists $\lambda\in[0,1]$ such that for each competition $(N,\mathcal{R},E)$ and each competitor $i\in N$, we have\footnote{Here, we define $\lambda^0=1$ for each $\lambda\in[0,1]$.}
\begin{equation*}
    \varphi_i(N,\mathcal{R},E)=\frac{\lambda^{\mathcal{R}(i)-1}}{\sum_{k=1}^{|N|}\lambda^{k-1}}E.
\end{equation*}
\end{property}

Note that the ED rule is a geometric rule with $\lambda=1$. The WTA rule is a geometric rule with $\lambda=0$. The prize allocation in the final stage of the poker tournament presented in Table~\ref{table:golfpoker}(a) is a geometric rule with $\lambda=0.713$. The prize allocation paths of the ED rule, the WTA rule, and the geometric rule $\varphi$ in the poker tournament are illustrated as follows.
\begin{center}\begin{tikzpicture}
\draw[->] (0,0) -- (5,0) node[anchor=north] {$\varphi_1$};
\draw	(0,0) node[anchor=north] {0}
        (1,0) node[anchor=north] {1}
        (2,0) node[anchor=north] {2}
        (3,0) node[anchor=north] {3}
		(4,0) node[anchor=north] {4};
\draw[->] (0,0) -- (0,5) node[anchor=east] {$\varphi_2$};
\draw	(0,1) node[anchor=east] {1}
		(0,2) node[anchor=east] {2}
		(0,3) node[anchor=east] {3}
        (0,4) node[anchor=east] {4};
\draw[dashed] (1,0) -- (0,1);
\draw[dashed] (2,0) -- (0,2);
\draw[dashed] (3,0) -- (0,3);
\draw[dashed] (4,0) -- (0,4);
\draw[dashed] (5,0) -- (0,5);
\draw[dashed] (5,1) -- (1,5);
\draw[dashed] (5,2) -- (2,5);
\draw[dashed] (5,3) -- (3,5);
\draw[dashed] (5,4) -- (4,5);
\draw[very thick] (0,0) -- (5,5);
\draw (3.2,4.5) node {${\rm ED}(N,\mathcal{R},E)$};
\draw[very thick] (0,0) -- (5,0);
\draw (3.8,0.5) node {${\rm WTA}(N,\mathcal{R},E)$};
\draw[very thick] (0,0) -- (5,3.565);
\draw (4,1.9) node {$\varphi(N,\mathcal{R},E)$};
\end{tikzpicture}\end{center}

\begin{corollary}\label{cor:geo}\ \\
A prize allocation rule satisfies anonymity, order preservation, endowment additivity (scale invariance), and local consistency if and only if it is a geometric rule.
\end{corollary}

\subsection{Top consistent prize allocation rules}\label{sec:topcons}

Theorem~\ref{theorem:localcons} shows that many rules satisfy anonymity, order preservation, winner strict endowment monotonicity, and local consistency. Strengthening order preservation or winner strict endowment monotonicity leads to direct restrictions on the representation of the single-parametric rules. Yet, there are still several rules observed in practice that do not satisfy local consistency, such as the golf tournament in Table~\ref{table:golfpoker}. To describe these rules, it may be desirable to weaken consistency even further to \emph{top consistency}, which requires only consistent allocations for reduced competitions where only competitors with the highest positions are involved. In other words, top consistency requires invariance for the competitors in the top segment of the ranking when they are put in a separate league. In competitions where competitors are eliminated sequentially, top consistency can be interpreted as invariance under prize allocation at the moment of elimination.

\begin{property}{Axiom: top consistency}\ \\
For each competition $(N,\mathcal{R},E)$, each subset of competitors $S\subseteq N$ such that $\mathcal{R}(i)\leq|S|$ for each competitor $i\in S$, and each competitor $i\in S$, we have
\begin{equation*}
    \varphi_i(N,\mathcal{R},E)=\varphi_i\left(S,\mathcal{R}_S,\sum_{j\in S}\varphi_j(N,\mathcal{R},E)\right).
\end{equation*}
\end{property}

Needless to say, all single-parametric rules satisfy anonymity, order preservation, winner strict endowment monotonicity, and top consistency, but these rules are not the only ones. The following example provides another rule satisfying these properties.

\begin{example}\label{ex:hyperarithmetic}\ \\
{\rm Let a prize allocation rule be defined by allocating the prize endowment in the following way. Up to the first two dollars are allocated to the competitor with the highest position. The surplus is divided equally among the competitors with the highest position and the second highest position until they are allocated three and one dollar, respectively. Then the surplus is divided equally among the competitors with the three highest positions until they are allocated four, two, and one dollar, respectively. This procedure continues until the competitor with the lowest position is allocated one dollar. If there is still money left, this is equally divided among the competitors. Table~\ref{table:hyperarithmetic} illustrates how a prize endowment from one to eight dollars is allocated among two, three, and four competitors. This rule also satisfies anonymity, order preservation, winner strict endowment monotonicity, and top consistency.
\closeex}
\end{example}

\begin{table}[ht!]
\caption{Eight dollars allocated among two, three, and four competitors in Example~\ref{ex:hyperarithmetic}}\label{table:hyperarithmetic}
{\small
\begin{center}
\begin{tabular}{cc|c}
\toprule
$\varphi_1$&	$\varphi_2$& $E$\\
\hline
1&	0&	1\\
2&	0&	2\\
5/2&    1/2&	3\\
3&	1&	4\\
7/2&	3/2&	5\\
4&	2&	6\\
9/2& 5/2& 7\\
5& 3& 8\\
\bottomrule
\end{tabular}
\quad
\begin{tabular}{ccc|c}
\toprule
$\varphi_1$&	$\varphi_2$&	$\varphi_3$&    $E$\\
\hline
1&	0&	0&	1\\
2&	0&	0&	2\\
5/2&	1/2&	0&	3\\
3&	1&	0&	4\\
10/3&	4/3&	1/3&	5\\
11/3&	5/3&	2/3&	6\\
4& 2& 1& 7\\
13/3& 7/3& 4/3& 8\\
\bottomrule
\end{tabular}
\quad
\begin{tabular}{cccc|c}
\toprule
$\varphi_1$&	$\varphi_2$&	$\varphi_3$&	$\varphi_4$& $E$\\
\hline
1&	0&	0&	0&	1\\
2&	0&	0&	0&	2\\
5/2&	1/2&	0&	0&	3\\
3&	1&	0&	0&	4\\
10/3&	4/3&	1/3&	0&	5\\
11/3&	5/3&	2/3&	0&	6\\
4& 2& 1& 0& 7\\
17/4& 9/4& 5/4& 1/4& 8\\
\bottomrule
\end{tabular}
\end{center}}
\end{table}

Each prize allocation rule satisfying anonymity, order preservation, winner strict endowment monotonicity, and top consistency admits a compact indirect description. For each such rule there exist continuous and non-decreasing functions $f_1,f_2,\ldots:\mathbb{R}_+\to\mathbb{R}_+$ with $f_1(x)=x$ for each $x$ and $f_{k+1}(x)\leq f_k(x)$ for each $k$ such that the assigned prize allocation for each competition can be described in the following way. The competitor with the highest position is allocated a prize of $f_1(x)=x$. The competitor with the second highest position is allocated a prize of $f_2(x)$. Generally, the prize for position $k$ is $f_k(x)$. In total, the full prize endowment is allocated among the competitors. We call such prize allocation rules \emph{parametric rules}.

\begin{property}{Parametric rules}\ \\
There exist continuous and non-decreasing functions $f_1,f_2,\ldots:\mathbb{R}_+\to\mathbb{R}_+$ with $f_1(x)=x$ for each $x$ and $f_{k+1}(x)\leq f_k(x)$ for each $k$ such that for each competition $(N,\mathcal{R},E)$ and each competitor $i\in N$, we have
\begin{equation*}
    \varphi_i(N,\mathcal{R},E)=f_{\mathcal{R}(i)}(x),
\end{equation*}
where $x\in\mathbb{R}_+$ is such that $\sum_{k=1}^{|N|}f_k(x)=E$.
\end{property}

Note that each single-parametric rule with representation $f$ is a parametric rule with $f_k=f^{(k-1)}$ for each $k$. In particular, the ED rule is a parametric rule with $f_k(x)=x$ for each $k$. The WTA rule is a parametric rule with $f_1(x)=x$ and $f_k(x)=0$ for each $k\geq2$. A Winner-Takes-Surplus rule (WTS) with representation $a$ is a parametric rule with $f_1(x)=x$ and $f_k(x)=\min\{a,x\}$ for each $k\geq2$. A geometric rule with representation $\lambda$ is a parametric rule with $f_k(x)=\lambda^{k-1}x$ for each $k$. The rule from Example~\ref{ex:hyperarithmetic} is a parametric rule with $f_1(x)=x$ and $f_k(x)=\max\{0,x-k\}$ for each $k\geq2$.

\bigskip

\begin{theorem}\label{theorem:topcons}\ \\
A prize allocation rule satisfies anonymity, order preservation, winner strict endowment monotonicity, and top consistency if and only if it is a parametric rule.
\end{theorem}

\bigskip

Subfamilies of parametric rules are again directly obtained if order preservation is strengthened to strict order preservation or winner-loser strict order preservation, or winner strict endowment monotonicity is strengthened to strict endowment monotonicity.

\begin{corollary}\label{cor:topcons}\ \\
Let $\varphi$ be a parametric rule with representation $f_1,f_2,\ldots$. Then the following statements hold.
\begin{enumerate}[(i)]
    \item $\varphi$ satisfies winner-loser strict order preservation if and only if $f_2(x)<x$ for each $x>0$;
    \item $\varphi$ satisfies strict order preservation if and only if $f_{k+1}(x)<f_k(x)$ for each $k$ and each $x>0$;
    \item $\varphi$ satisfies strict endowment monotonicity if and only if $f_k(x)$ is increasing for each $k$.
\end{enumerate}
\end{corollary}

If winner strict endowment monotonicity is strengthened to endowment additivity (scale invariance), then the family of widely used \emph{proportional rules} is obtained. For each proportional rule, there exist $\lambda_1,\lambda_2,\ldots\in\mathbb{R}_+$ with $\lambda_1>0$ and $\lambda_{k+1}\leq\lambda_k$ for each $k$ such that for each competition the prize for position $k$ is proportional to $\lambda_k$. In total, the full prize endowment is allocated.

\begin{property}{Proportional rules}\ \\
There exist $\lambda_1,\lambda_2,\ldots\in\mathbb{R}_+$ with $\lambda_1>0$ and $\lambda_{k+1}\leq\lambda_k$ for each $k$ such that for each competition $(N,\mathcal{R},E)$ and each competitor $i\in N$, we have
\begin{equation*}
    \varphi_i(N,\mathcal{R},E)=\frac{\lambda_{\mathcal{R}(i)}}{\sum_{k=1}^{|N|}\lambda_k}E.
\end{equation*}
\end{property}

Note that each geometric rule with representation $\lambda$ is a proportional rule with $\lambda_k=\lambda^{k-1}$ for each $k$. In particular, the ED rule is a proportional rule with $\lambda_k=1$ for each $k$. The WTA rule is a proportional rule with $\lambda_1=1$ and $\lambda_k=0$ for each $k\geq2$. The prize allocation rule in the golf tournament of Table~\ref{table:golfpoker}(b) is a proportional rule with $\lambda_1=18.0$, $\lambda_2=10.9$, $\lambda_3=6.9$, and so on.

\begin{corollary}\label{cor:topprop}\ \\
A prize allocation rule satisfies anonymity, order preservation, endowment additivity (scale invariance), and top consistency if and only if it is a proportional rule.
\end{corollary}


\section{Concluding remarks}\label{sec:conrem}

In this paper, we initiate an axiomatic approach to study prize allocation rules in rank-order competitions. We introduce a model in which the competitors, their final ranking, and the prize endowment are the primitives and we axiomatically characterize three families of prize allocation rules: interval rules (Theorem~\ref{theorem:cons}), single-parametric rules (Theorem~\ref{theorem:localcons}), and parametric rules (Theorem~\ref{theorem:topcons}). In the Appendix, we show that the axioms used are logically independent. Moreover, we have obtained several subfamilies: Winner-Takes-Surplus rules (Corollary~\ref{cor:WTS}), geometric rules (Corollary~\ref{cor:geo}), and proportional rules (Corollary~\ref{cor:topprop}). Each of these families and subfamilies includes the Equal Division rule and the Winner-Takes-All rule. The relations of the families are presented in the picture below.

\begin{center}
\includegraphics[scale=0.5]{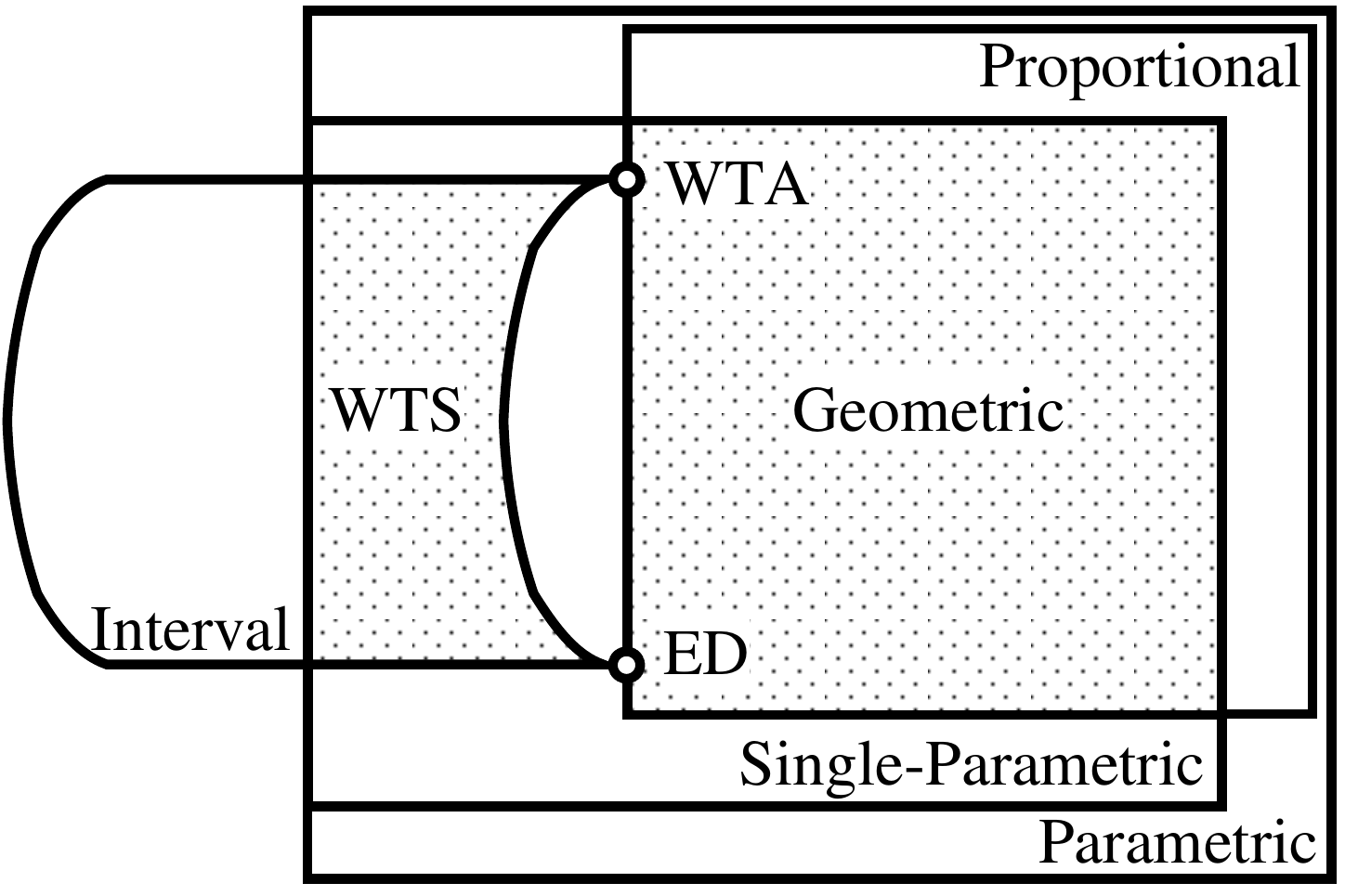}
\end{center}

The practical relevance of our paper is that if an organizer wants to choose a prize allocation rule and accepts some set of axioms, then the problem is reduced to choosing a single rule from the corresponding family. Since the families are broad enough, the organizer can choose the rule that maximizes some desirable goal, such as the total effort of competitors or the participation of talented competitors. This problem could be studied in future research using an optimization approach.

Our axiomatic framework can be further developed in many directions. For single rank-order tournaments, a natural direction is the study of other desirable axioms. In situations where more data is available, we can take more details into account for prize allocations. One possibility is to replace the ordinal ranking in our model by a cardinal ranking, e.g., using finish times, scores, or the volume of sales. Another possibility is to consider other competition types, such as knockout tournaments, round-robin tournaments, and multiple rank-order tournaments.

Multiple rank-order tournaments, consisting of a series of single rank-order tournaments, are interesting for the following reasons. In many real series, a competitor gets a prize and a number of points in each single tournament. Then the total sum of points determines the total ranking and the bonuses for the entire series. A straightforward question is how to jointly choose a points system, a prize structure for each single tournament, and a bonus structure for the entire series. In particular, since Corollary~\ref{cor:geo} calls for geometric prize sequences and \citet{KondratevIanovskiNesterov19} justify geometric point sequences, should we choose the same parameter for both geometric sequences?

Another open question is how to apply for competitions the rules developed for ranking, voting, or budget allocation. For instance, \citet{Kreweras65} and \citet{Fishburn84lotteries} developed a probabilistic voting rule known as maximal lotteries. \citet{BrandlBrandtSeedig16} noted that ‘the lotteries returned by probabilistic social choice functions do not necessarily have to be interpreted as probability distributions. They can, for instance, also be seen as fractional allocations of divisible objects such as time shares or monetary budgets.’ \citet{Airiau19} argued that ‘the maximal lotteries rule, while attractive according to consistency axioms, spends the entire budget on the Condorcet winner if it exists. This is often undesirable in a budgeting context.’ We can conclude from these arguments that any application of well-known ranking, voting, or allocation rules must be re-motivated and re-justified in the context of competitions.

\section*{Appendix}

\renewcommand\thelemma{\ref{lemma:econt}}
\begin{lemma}\ \\
If a prize allocation rule satisfies endowment monotonicity, then it satisfies endowment continuity.
\begin{proof}
Let $\varphi$ be a prize allocation rule satisfying endowment monotonicity. Let $(N,\mathcal{R},E)$ and $(N,\mathcal{R},E')$ be two competitions. Let $i\in N$. By endowment monotonicity,
\begin{alignat*}1
    |E-E'| &= |\sum_{j\in N}\varphi_j(N,\mathcal{R},E)-\sum_{j\in N}\varphi_j(N,\mathcal{R},E')|
\\  &= |\sum_{j\in N}(\varphi_j(N,\mathcal{R},E)-\varphi_j(N,\mathcal{R},E'))|
\\  &= \sum_{j\in N}|\varphi_j(N,\mathcal{R},E)-\varphi_j(N,\mathcal{R},E')|
\\  &\geq |\varphi_i(N,\mathcal{R},E)-\varphi_i(N,\mathcal{R},E')|.
\end{alignat*}
This means that $\varphi_i(N,\mathcal{R},E)\to\varphi_i(N,\mathcal{R},E')$ if $E\to E'$. Hence, $\varphi$ satisfies endowment continuity.
\end{proof}
\end{lemma}

\renewcommand\thelemma{\ref{lemma:anonymity}}
\begin{lemma}\ \\
If a prize allocation rule satisfies order preservation, endowment monotonicity, and consistency, then it satisfies anonymity.
\begin{proof}
Let $\varphi$ be a prize allocation rule satisfying order preservation, endowment monotonicity, and consistency. By Lemma~\ref{lemma:econt}, $\varphi$ satisfies endowment continuity. 

Suppose for the sake of contradiction that $\varphi$ does not satisfy anonymity. Then there exist two competitions $(N,\mathcal{R},E)$ and $(N',\mathcal{R}',E)$ with equal numbers of competitors $|N|=|N'|$, and two competitors $i_1\in N$ and $j_1\in N'$ with equal positions $\mathcal{R}(i_1)=\mathcal{R}'(j_1)$ such that $\varphi_{i_1}(N,\mathcal{R},E)<\varphi_{j_1}(N',\mathcal{R}',E)$. Since 
\begin{equation*}
    \sum_{k\in N}\varphi_k(N,\mathcal{R},E)=E=\sum_{k\in N'}\varphi_k(N',\mathcal{R}',E), 
\end{equation*}
there exist two competitors $i_2\in N$ and $j_2\in N'$ with equal position $\mathcal{R}(i_2)=\mathcal{R}'(j_2)$ such that $\varphi_{i_2}(N,\mathcal{R},E)>\varphi_{j_2}(N',\mathcal{R}',E)$. Suppose without loss of generality that $\mathcal{R}'(j_1)=\mathcal{R}(i_1)<\mathcal{R}(i_2)=\mathcal{R}'(j_2)$. Denote $x=\varphi(N,\mathcal{R},E)$ and $y=\varphi(N',\mathcal{R}',E)$. By consistency,
\begin{equation*}
    (x_{i_1},x_{i_2})=\varphi(\{i_1,i_2\},\mathcal{R}_{\{i_1,i_2\}},x_{i_1}+x_{i_2})
\end{equation*}
and
\begin{equation*}
    (y_{j_1},y_{j_2})=\varphi(\{j_1,j_2\},\mathcal{R}_{\{j_1,j_2\}}',y_{j_1}+y_{j_2}).
\end{equation*}
This is illustrated in the following way.
\begin{equation}\label{tables}
\begin{tabular}{l|l||l}
$N$ & $\mathcal{R}$ & $\varphi$ \\ \hline
$i_1$ & $1$ & $x_{i_1}$ \\
$i_2$ & $2$ & $x_{i_2}$
\end{tabular}
\quad\quad
\begin{tabular}{l|l||l}
$N$ & $\mathcal{R}$ & $\varphi$ \\ \hline
$j_1$ & $1$ & $y_{j_1}$ \\
$j_2$ & $2$ & $y_{j_2}$
\end{tabular}
\end{equation}
By order preservation, $y_{j_2}<x_{i_2}\leq x_{i_1}<y_{j_1}$. One of the following six cases hold.

\begin{case}{$i_1=j_2$ and $i_2=j_1$}\label{case}\ \\
{\rm By (\ref{tables}),
\begin{equation*}
\begin{tabular}{l|l||l}
$N$ & $\mathcal{R}$ & $\varphi$ \\ \hline
$i_1=j_2$ & $1$ & $x_{i_1}$ \\
$i_2=j_1$ & $2$ & $x_{i_2}$
\end{tabular}
\quad\quad
\begin{tabular}{l|l||l}
$N$ & $\mathcal{R}$ & $\varphi$ \\ \hline
$j_1$ & $1$ & $y_{j_1}$ \\
$j_2$ & $2$ & $y_{j_2}$
\end{tabular}
\end{equation*}
Then there exists another competitor $k\in U\setminus\{i_1,i_2\}$. By order preservation, endowment continuity, and consistency, there exist prize endowments $E'$ and $E''$ such that
\begin{equation*}
\begin{tabular}{l|l||l}
$N$ & $\mathcal{R}$ & $\varphi$ \\ \hline
$i_1$ & $1$ & $x_{i_1}$ \\
$i_2$ & $2$ & $x_{i_2}$ \\
$k$ & $3$ & $E'-x_{i_1}-x_{i_2}$
\end{tabular}
\quad\quad
\begin{tabular}{l|l||l}
$N$ & $\mathcal{R}$ & $\varphi$ \\ \hline
$i_1$ & $1$ & $x_{i_1}$ \\
$k$ & $2$ & $E''-x_{i_1}-x_{i_2}$ \\
$i_2$ & $3$ & $x_{i_2}$
\end{tabular}
\end{equation*}
By order preservation, $E'-x_{i_1}-x_{i_2}\leq x_{i_2}\leq E''-x_{i_1}-x_{i_2}$. By consistency,
\begin{equation*}
\begin{tabular}{l|l||l}
$N$ & $\mathcal{R}$ & $\varphi$ \\ \hline
$i_1$ & $1$ & $x_{i_1}$ \\
$k$ & $2$ & $E'-x_{i_1}-x_{i_2}$
\end{tabular}
\quad\quad
\begin{tabular}{l|l||l}
$N$ & $\mathcal{R}$ & $\varphi$ \\ \hline
$i_1$ & $1$ & $x_{i_1}$ \\
$k$ & $2$ & $E''-x_{i_1}-x_{i_2}$ \\
\end{tabular}
\end{equation*}
By endowment monotonicity,
\begin{equation*}
\begin{tabular}{l|l||l}
$N$ & $\mathcal{R}$ & $\varphi$ \\ \hline
$i_1$ & $1$ & $x_{i_1}$ \\
$k$ & $2$ & $x_{i_2}$
\end{tabular}
\end{equation*}
By order preservation, endowment continuity, consistency, and (\ref{tables}), there exist prize endowments $E'''$ and $E''''$ such that
\begin{equation*}
\begin{tabular}{l|l||l}
$N$ & $\mathcal{R}$ & $\varphi$ \\ \hline
$i_1$ & $1$ & $x_{i_1}$ \\
$i_2$ & $2$ & $E'''-x_{i_1}-x_{i_2}$ \\
$k$ & $3$ & $x_{i_2}$
\end{tabular}
\quad\quad
\begin{tabular}{l|l||l}
$N$ & $\mathcal{R}$ & $\varphi$ \\ \hline
$j_1$ & $1$ & $y_{j_1}$ \\
$j_2$ & $2$ & $y_{j_2}$ \\
$k$ & $3$ & $E''''-y_{j_1}-y_{j_2}$
\end{tabular}
\end{equation*}
By order preservation, $E'''-x_{i_1}-x_{i_2}\leq x_{i_1}$ and $E''''-y_{j_1}-y_{j_2}\leq y_{j_2}$. By consistency,
\begin{equation*}
\begin{tabular}{l|l||l}
$N$ & $\mathcal{R}$ & $\varphi$ \\ \hline
$i_2$ & $1$ & $E'''-x_{i_1}-x_{i_2}$ \\
$k$ & $2$ & $x_{i_2}$
\end{tabular}
\quad\quad
\begin{tabular}{l|l||l}
$N$ & $\mathcal{R}$ & $\varphi$ \\ \hline
$i_2=j_1$ & $1$ & $y_{j_1}$ \\
$k$ & $2$ & $E''''-y_{j_1}-y_{j_2}$
\end{tabular}
\end{equation*}
Then $E'''-x_{i_1}-x_{i_2}\leq x_{i_1}<y_{j_1}$ and $x_{i_2}>y_{j_2}\geq E''''-y_{j_1}-y_{j_2}$. This contradicts endowment monotonicity.}
\end{case}

\begin{case}{$i_1=j_2$ and $i_2\neq j_1$}\ \\
{\rm By order preservation, endowment continuity, consistency, and (\ref{tables}), there exists prize endowment $E'$ such that
\begin{equation*}
\begin{tabular}{l|l||l}
$N$ & $\mathcal{R}$ & $\varphi$ \\ \hline
$i_1$ & $1$ & $x_{i_1}$ \\
$j_1$ & $2$ & $E'-x_{i_1}-x_{i_2}$ \\
$i_2$ & $3$ & $x_{i_2}$
\end{tabular}
\end{equation*}
By order preservation, $E'-x_{i_1}-x_{i_2}\geq x_{i_2}$. By consistency,
\begin{equation*}
\begin{tabular}{l|l||l}
$N$ & $\mathcal{R}$ & $\varphi$ \\ \hline
$i_1=j_2$ & $1$ & $x_{i_1}$ \\
$j_1$ & $2$ & $E'-x_{i_1}-x_{i_2}$
\end{tabular}
\end{equation*}
Then $x_{i_1}<y_{j_1}$ and $E'-x_{i_1}-x_{i_2}\geq x_{i_2}>y_{j_2}$. By (\ref{tables}), a contradiction follows from Case~\ref{case}.}
\end{case}

\begin{case}{$i_1\neq j_2$ and $i_2=j_1$}\ \\
{\rm By order preservation, endowment continuity, consistency, and (\ref{tables}), there exists a prize endowment $E'$ such that
\begin{equation*}
\begin{tabular}{l|l||l}
$N$ & $\mathcal{R}$ & $\varphi$ \\ \hline
$i_1$ & $1$ & $x_{i_1}$ \\
$j_2$ & $2$ & $E'-x_{i_1}-x_{i_2}$ \\
$i_2$ & $3$ & $x_{i_2}$
\end{tabular}
\end{equation*}
By order preservation, $E'-x_{i_1}-x_{i_2}\leq x_{i_1}$. By consistency,
\begin{equation*}
\begin{tabular}{l|l||l}
$N$ & $\mathcal{R}$ & $\varphi$ \\ \hline
$j_2$ & $1$ & $E'-x_{i_1}-x_{i_2}$ \\
$i_2=j_1$ & $2$ & $x_{i_2}$
\end{tabular}
\end{equation*}
Then $E'-x_{i_1}-x_{i_2}\leq x_{i_1}<y_{j_1}$ and $x_{i_2}>y_{j_2}$. By (\ref{tables}), a contradiction follows from Case~\ref{case}.}
\end{case}

\begin{case}{$i_1=j_1$ and $i_2\neq j_2$}\ \\
{\rm By order preservation, endowment continuity, consistency, and (\ref{tables}), there exists a prize endowment $E'$ such that
\begin{equation*}
\begin{tabular}{l|l||l}
$N$ & $\mathcal{R}$ & $\varphi$ \\ \hline
$i_1$ & $1$ & $x_{i_1}$ \\
$j_2$ & $2$ & $E'-x_{i_1}-x_{i_2}$ \\
$i_2$ & $3$ & $x_{i_2}$
\end{tabular}
\end{equation*}
By order preservation, $E'-x_{i_1}-x_{i_2}\geq x_{i_2}$. By consistency,
\begin{equation*}
\begin{tabular}{l|l||l}
$N$ & $\mathcal{R}$ & $\varphi$ \\ \hline
$i_1=j_1$ & $1$ & $x_{i_1}$ \\
$j_2$ & $2$ & $E'-x_{i_1}-x_{i_2}$
\end{tabular}
\end{equation*}
Then $x_{i_1}<y_{j_1}$ and $E'-x_{i_1}-x_{i_2}\geq x_{i_2}>y_{j_2}$. By (\ref{tables}), this contradicts endowment monotonicity.}
\end{case}

\begin{case}{$i_1\neq j_1$ and $i_2=j_2$}\ \\
{\rm By order preservation, endowment continuity, consistency, and (\ref{tables}), there exists a prize endowment $E'$ such that
\begin{equation*}
\begin{tabular}{l|l||l}
$N$ & $\mathcal{R}$ & $\varphi$ \\ \hline
$i_1$ & $1$ & $x_{i_1}$ \\
$j_1$ & $2$ & $E'-x_{i_1}-x_{i_2}$ \\
$i_2$ & $3$ & $x_{i_2}$
\end{tabular}
\end{equation*}
By order preservation, $E'-x_{i_1}-x_{i_2}\leq x_{i_1}$. By consistency,
\begin{equation*}
\begin{tabular}{l|l||l}
$N$ & $\mathcal{R}$ & $\varphi$ \\ \hline
$j_1$ & $1$ & $E'-x_{i_1}-x_{i_2}$ \\
$i_2=j_2$ & $2$ & $x_{i_2}$
\end{tabular}
\end{equation*}
Then $E'-x_{i_1}-x_{i_2}\leq x_{i_1}<y_{j_1}$ and $x_{i_2}>y_{j_2}$. By (\ref{tables}), this contradicts endowment monotonicity.}
\end{case}

\begin{case}{$i_1\neq j_1\neq i_2$ and $i_1\neq j_2\neq i_2$}\ \\
{\rm By order preservation, endowment continuity, consistency, and (\ref{tables}), there exist $z_{j_1}$ and $z_{j_2}$ such that
\begin{equation*}
\begin{tabular}{l|l||l}
$N$ & $\mathcal{R}$ & $\varphi$ \\ \hline
$i_1$ & $1$ & $x_{i_1}$ \\
$j_1$ & $2$ & $z_{j_1}$ \\
$j_2$ & $3$ & $z_{j_2}$ \\
$i_2$ & $4$ & $x_{i_2}$
\end{tabular}
\end{equation*}
By order preservation, $z_{j_1}\leq x_{i_1}$ and $z_{j_2}\geq x_{i_2}$. By consistency,
\begin{equation*}
\begin{tabular}{l|l||l}
$N$ & $\mathcal{R}$ & $\varphi$ \\ \hline
$j_1$ & $1$ & $z_{j_1}$ \\
$j_2$ & $2$ & $z_{j_2}$
\end{tabular}
\end{equation*}
Then $z_{j_1}\leq x_{i_1}<y_{j_1}$ and $z_{j_2}\geq x_{i_2}>y_{j_2}$. By (\ref{tables}), this contradicts endowment monotonicity.}
\end{case}
\end{proof}
\end{lemma}

\renewcommand\thelemma{\ref{lemma:cons}}
\begin{lemma}\ \\
If two prize allocation rules satisfying endowment monotonicity and consistency coincide for each competition with two competitors, then the two rules coincide for each competition with an arbitrary number of competitors.
\begin{proof}
Let $\varphi$ and $\varphi'$ be two prize allocation rules satisfying endowment monotonicity and consistency such that $\varphi(N,\mathcal{R},E)=\varphi'(N,\mathcal{R},E)$ for each competition with two competitors. Suppose for the sake of contradiction that there exists a competition $(N,\mathcal{R},E)$ such that $\varphi(N,\mathcal{R},E)\neq\varphi'(N,\mathcal{R},E)$. Since
\begin{equation*}
    \sum_{i\in N}\varphi_i(N,\mathcal{R},E)=E=\sum_{i\in N}\varphi_i'(N,\mathcal{R},E), 
\end{equation*}
there exist two competitors $i\in N$ and $j\in N$ such that
\begin{equation}\label{contradiction}
    \varphi_i(N,\mathcal{R},E)<\varphi_i'(N,\mathcal{R},E) \mbox{ and } \varphi_j(N,\mathcal{R},E)>\varphi_j'(N,\mathcal{R},E). 
\end{equation}
By consistency,
\begin{equation*}
    (\varphi_i(N,\mathcal{R},E),\varphi_j(N,\mathcal{R},E))=\varphi(\{i,j\},\mathcal{R}_{\{i,j\}},\varphi_i(N,\mathcal{R},E)+\varphi_j(N,\mathcal{R},E))
\end{equation*}
and
\begin{alignat*}2
    (\varphi_i'(N,\mathcal{R},E),\varphi_j'(N,\mathcal{R},E)) &= \varphi'(\{i,j\},\mathcal{R}_{\{i,j\}},\varphi_i'(N,\mathcal{R},E)+\varphi_j'(N,\mathcal{R},E))
\\  &= \varphi(\{i,j\},\mathcal{R}_{\{i,j\}},\varphi_i'(N,\mathcal{R},E)+\varphi_j'(N,\mathcal{R},E)).
\end{alignat*}
By endowment monotonicity,
\begin{equation*}
    \varphi_i(N,\mathcal{R},E)\leq\varphi_i'(N,\mathcal{R},E)\mbox{ and }\varphi_j(N,\mathcal{R},E)\leq\varphi_j'(N,\mathcal{R},E),
\end{equation*}
or
\begin{equation*}
    \varphi_i(N,\mathcal{R},E)\geq\varphi_i'(N,\mathcal{R},E)\mbox{ and }\varphi_j(N,\mathcal{R},E)\geq\varphi_j'(N,\mathcal{R},E).
\end{equation*}
This contradicts (\ref{contradiction}). Hence, $\varphi(N,\mathcal{R},E)=\varphi'(N,\mathcal{R},E)$ for each competition $(N,\mathcal{R},E)$.
\end{proof}
\end{lemma}

\bigskip

\renewcommand\thetheorem{\ref{theorem:cons}}
\begin{theorem}\ \\
A prize allocation rule satisfies order preservation, endowment monotonicity, and consistency if and only if it is an interval rule.
\begin{proof}
It is readily checked that each interval rule satisfies order preservation, endowment monotonicity, and consistency. 

Let $\varphi$ be a prize allocation rule satisfying order preservation, endowment monotonicity, and consistency. By Lemma~\ref{lemma:econt}, $\varphi$ satisfies endowment continuity. By Lemma~\ref{lemma:anonymity}, $\varphi$ satisfies anonymity. By Lemma~\ref{lemma:cons}, we only need to show that $\varphi$ is an interval rule for each competition with two competitors, since each such rule has a unique consistent extension to competitions with more competitors. Let $N\subseteq U$ with $|N|=2$ be two competitors and let $\mathcal{R}$ be a ranking. Denote $N=\{1,2\}$ such that $\mathcal{R}(1)=1$ and $\mathcal{R}(2)=2$. The proof consists of two steps.

\bigskip

\noindent \textbf{Step~1.}
For each prize endowment $E$, if $E=x_1+x_2$ such that
\begin{equation}\label{x1x2}
    \varphi(N,\mathcal{R},x_1+x_2)=(x_1,x_2),
\end{equation}
then
\begin{equation}\label{OR}
    \varphi(N,\mathcal{R},x_1+x_1)=(x_1,x_1)\ \text{or}\ \varphi(N,\mathcal{R},x_2+x_2)=(x_2,x_2).
\end{equation}

\noindent \emph{Proof of Step~1.}
Let $E$ be a prize endowment and denote $\varphi(N,\mathcal{R},E)=(x_1,x_2)$. Then $E=x_1+x_2$ and (\ref{x1x2}) holds. By order preservation, $x_1\geq x_2$. If $x_1=x_2$, then (\ref{OR}) follows immediately from (\ref{x1x2}).

Suppose that $x_1>x_2$. Let $N'=\{1,2,3\}$ and let $\mathcal{R}'$ be a ranking of $N'$ such that $\mathcal{R}'(1)=1$, $\mathcal{R}'(2)=2$, and $\mathcal{R}'(3)=3$. By order preservation and endowment continuity, there exists a prize endowment $E'$ such that $\varphi_1(N',\mathcal{R}',E')+\varphi_3(N',\mathcal{R}',E')=x_1+x_2$. By anonymity, consistency, and (\ref{x1x2}), 
\begin{equation}\label{E'}
    \varphi(N',\mathcal{R}',E')=(x_1,E'-x_1-x_2,x_2).
\end{equation}
By order preservation, $x_1+x_2+x_2\leq E'\leq x_1+x_1+x_2$. By anonymity and consistency,
\begin{alignat}2
&\varphi(N,\mathcal{R},E'-x_1) &&= (E'-x_1-x_2,x_2)\label{E'x1}
\\ \text{and}\ &\varphi(N,\mathcal{R},E'-x_2) &&= (x_1,E'-x_1-x_2).\label{E'x2}
\end{alignat}
If $E'=x_1+x_2+x_2$, then (\ref{OR}) follows immediately from (\ref{E'x1}). If $E'=x_1+x_1+x_2$, then (\ref{OR}) follows immediately from (\ref{E'x2}).

Suppose that $x_1+x_2+x_2<E'<x_1+x_1+x_2$. Then $E'-x_1<x_1+x_2<E'-x_2$. By endowment monotonicity, (\ref{x1x2}), (\ref{E'x1}), and (\ref{E'x2}),
\begin{equation}\label{cases}
    \varphi(N,\mathcal{R},E'')=
    \begin{cases}
        (E''-x_2,x_2) & \text{if $E'-x_1\leq E''\leq x_1+x_2$;}
    \\  (x_1,E''-x_1) & \text{if $x_1+x_2\leq E''\leq E'-x_2$.}
    \end{cases}
\end{equation}
Denote $(y_1,y_2,y_3)=\varphi(N',\mathcal{R}',x_1+x_2+x_2)$ and $(z_1,z_2,z_3)=\varphi(N',\mathcal{R}',x_1+x_1+x_2)$. By endowment monotonicity and (\ref{E'}),
\begin{alignat*}3
    y_1 &\leq  &&x_1  &&\leq z_1,
\\ y_2 &\leq E'- &&x_1 -x_2 &&\leq z_2,
\\ y_3 &\leq &&x_2 &&\leq z_3.
\end{alignat*}
Then $y_1+y_2\leq E'-x_2$ and $E'-x_1 \leq z_2+z_3$.
Since $y_3\leq x_2$ and $y_1+y_2+y_3=x_1+x_2+x_2$, we have $x_1+x_2 \leq x_1+x_2+x_2-y_3 = y_1+y_2$. Since $x_1 \leq z_1$ and $z_1+z_2+z_3=x_1+x_1+x_2$, we have $z_2+z_3=x_1+x_1+x_2-z_1\leq x_1+x_2$. This means that
\begin{alignat*}1
    &x_1+x_2\leq y_1+y_2\leq E'-x_2,
\\  \text{and}\ &E'-x_1\leq z_2+z_3\leq x_1+x_2.
\end{alignat*}
By anonymity, consistency, and (\ref{cases}),
\begin{alignat*}1
    &\varphi_1(N,\mathcal{R},y_1+y_3)=y_1=\varphi_1(N,\mathcal{R},y_1+y_2)=x_1,
\\  \text{and}\ &\varphi_2(N,\mathcal{R},z_1+z_3)=z_3=\varphi_2(N,\mathcal{R},z_2+z_3)=x_2.
\end{alignat*}
Since $y_1\leq z_1$ and $y_3\leq z_3$, we have $y_1+y_3\leq z_1+z_3$. By endowment monotonicity and (\ref{cases}), for each $E''$ such that $\varphi_1(N,\mathcal{R},E'')\geq x_1$, we have $E''\geq x_1+x_2$. Since $\varphi_1(N,\mathcal{R},y_1+y_3)=x_1$, we have $y_1+y_3\geq x_1+x_2$. By endowment monotonicity and (\ref{cases}), for each $E''$ such that $\varphi_2(N,\mathcal{R},E'')\leq x_2$, we have $E''\leq x_1+x_2$. Since $\varphi_2(N,\mathcal{R},z_1+z_3)=x_2$, we have $z_1+z_3\leq x_1+x_2$. Hence,
\begin{equation*}
    x_1+x_2\leq y_1+y_3\leq z_1+z_3\leq x_1+x_2.
\end{equation*}
Then $y_1+y_3=x_1+x_2$ and $z_1+z_3=x_1+x_2$. Since $y_1=x_1$ and $z_3=x_2$, we have $y_3=x_2$ and $z_1=x_1$. This means that $y=(x_1,x_2,x_2)$ and $z=(x_1,x_1,x_2)$. By anonymity and consistency, $\varphi(N,\mathcal{R},x_1+x_1)=(x_1,x_1)$ and $\varphi(N,\mathcal{R},x_2+x_2)=(x_2,x_2)$. Hence, (\ref{OR}) holds.

\bigskip

\noindent \textbf{Step~2.}
There exist disjoint intervals $(a_1,b_1),(a_2,b_2),\ldots$ with $a_1,a_2,\ldots\in\mathbb{R}_+$ and $b_1,b_2,\ldots\in\mathbb{R}_+\cup\{+\infty\}$ such that
\begin{equation}\label{description_2}
    \varphi(N,\mathcal{R},E)=
    \begin{cases}
        (E-a_k,a_k) & \text{if $a_k+a_k\leq E\leq b_k+a_k$};
    \\  (b_k,E-b_k) & \text{if $b_k+a_k\leq E\leq b_k+b_k$};
    \\  \frac{E}{2} & \text{otherwise}.
    \end{cases}
\end{equation}

\noindent \emph{Proof of Step~2.}
If $\varphi_1(N,\mathcal{R},E)=\varphi_2(N,\mathcal{R},E)$ for each prize endowment $E$, then (\ref{description_2}) follows immediately by defining $a_k=b_k=0$ for each $k$.

Let $E$ be a prize endowment with $\varphi_1(N,\mathcal{R},E)>\varphi_2(N,\mathcal{R},E)$. Denote $\varphi(N,\mathcal{R},E)=(x_1,x_2)$. Then $E=x_1+x_2$, $x_1>x_2$, and
\begin{equation*}
    \varphi(N,\mathcal{R},x_1+x_2)=(x_1,x_2).
\end{equation*}
By Step~1, $\varphi(N,\mathcal{R},x_1+x_1)=(x_1,x_1)$ or $\varphi(N,\mathcal{R},x_2+x_2)=(x_2,x_2)$.

Suppose that $\varphi(N,\mathcal{R},x_1+x_1)=(x_1,x_1)$. Define
\begin{equation}\label{ak}
\begin{alignedat}2
&b_E &&= x_1 
\\ \text{and}\ &a_E &&= \min_{E'\in\mathbb{R}_+}\{\varphi_2(N,\mathcal{R},E')\mid\varphi_1(N,\mathcal{R},E')=b_E\}.
\end{alignedat}
\end{equation}
Then $a_E<b_E$ since $a_E\leq x_2<x_1=b_E$. Moreover, $\varphi(N,\mathcal{R},b_E+a_E)=(b_E,a_E)$ and $\varphi(N,\mathcal{R},b_E+b_E)=(b_E,b_E)$. By endowment monotonicity,
\begin{equation}\label{part}
    \varphi(N,\mathcal{R},E')=(b_E,E'-b_E)\quad \text{if $b_E+a_E\leq E'\leq b_E+b_E$.}
\end{equation}
This also means that
\begin{equation}\label{extra}
    b_E+a_E<E'<b_E+b_E\quad \text{if $a_E<\varphi_2(N,\mathcal{R},E')<b_E$}.
\end{equation}
Let $E'$ be a prize endowment with $a_E+a_E<E'<b_E+a_E$. Denote $(y_1,y_2)=\varphi(N,\mathcal{R},E')$. By endowment monotonicity and (\ref{ak}), $y_1<b_E$ and $y_2\leq a_E$. Then $a_E<y_1$ since $a_E\leq a_E+a_E-y_2<E'-y_2=y_1$. By Step~1, $\varphi(N,\mathcal{R},y_1+y_1)=(y_1,y_1)$ or $\varphi(N,\mathcal{R},y_2+y_2)=(y_2,y_2)$. If $\varphi(N,\mathcal{R},y_1+y_1)=(y_1,y_1)$, then $a_E<y_1=\varphi_2(N,\mathcal{R},y_1+y_1)<b_E$, (\ref{extra}) implies $b_E+a_E<y_1+y_1<b_E+b_E$ and (\ref{part}) implies $y_1=\varphi_1(N,\mathcal{R},y_1+y_1)=b_E$, which is a contradiction. This means that $\varphi(N,\mathcal{R},\varphi_2(N,\mathcal{R},E')+\varphi_2(N,\mathcal{R},E'))=(\varphi_2(N,\mathcal{R},E'),\varphi_2(N,\mathcal{R},E'))$ for each prize endowment $E'$ with $a_E+a_E<E'<b_E+a_E$. Since $\varphi_2(N,\mathcal{R},b_E+a_E)=a_E$, we have $\varphi(N,\mathcal{R},a_E+a_E)=(a_E,a_E)$ by endowment continuity. By endowment monotonicity,
\begin{equation*}
    \varphi(N,\mathcal{R},E')=(E'-a_E,a_E)\quad \text{if $a_E+a_E\leq E'\leq b_E+a_E$.}
\end{equation*}

Suppose that $\varphi(N,\mathcal{R},x_2+x_2)=(x_2,x_2)$. Define
\begin{equation}\label{bk}
\begin{alignedat}2
&a_E &&= x_2
\\ \text{and}\ &b_E &&= \sup_{E'\in\mathbb{R}_+}\{\varphi_1(N,\mathcal{R},E')\mid\varphi_2(N,\mathcal{R},E')=a_E\}.
\end{alignedat}
\end{equation}
Then $a_E<b_E$ since $a_E= x_2<x_1\leq b_E$. Moreover, $\varphi(N,\mathcal{R},a_E+a_E)=(a_E,a_E)$ and $\varphi(N,\mathcal{R},b_E+a_E)=(b_E,a_E)$. By endowment monotonicity,
\begin{equation}\label{part2}
    \varphi(N,\mathcal{R},E')=(E'-a_E,a_E)\ \text{if $a_E+a_E\leq E'\leq b_E+a_E$.}
\end{equation}
This also means that
\begin{equation}\label{extra2}
    a_E+a_E<E'<b_E+a_E\ \text{if $a_E<\varphi_1(N,\mathcal{R},E')<b_E$}.
\end{equation}
Let $E'$ be a prize endowment with $b_E+a_E<E'<b_E+b_E$. Denote $(y_1,y_2)=\varphi(N,\mathcal{R},E')$. By endowment monotonicity and (\ref{bk}), $b_E\leq y_1$ and $a_E<y_2$. Then $y_2<b_E$ since $y_2=E'-y_1<b_E+b_E-y_1\leq b_E$. By Step~1, $\varphi(N,\mathcal{R},y_1+y_1)=(y_1,y_1)$ or $\varphi(N,\mathcal{R},y_2+y_2)=(y_2,y_2)$. If $\varphi(N,\mathcal{R},y_2+y_2)=(y_2,y_2)$, then $a_E<y_2=\varphi_1(N,\mathcal{R},y_2+y_2)<b_E$, (\ref{extra2}) implies $a_E+a_E<y_2+y_2<b_E+a_E$ and (\ref{part2}) implies $y_2=\varphi_2(N,\mathcal{R},y_2+y_2)=a_E$, which is a contradiction. This means that $\varphi(N,\mathcal{R},\varphi_1(N,\mathcal{R},E')+\varphi_1(N,\mathcal{R},E'))=(\varphi_1(N,\mathcal{R},E'),\varphi_1(N,\mathcal{R},E'))$ for each prize endowment $E'$ with $b_E+a_E<E'<b_E+b_E$. Since $\varphi_1(N,\mathcal{R},b_E+a_E)=b_E$, we have $\varphi(N,\mathcal{R},b_E+b_E)=(b_E,b_E)$ by endowment continuity.  By endowment monotonicity,
\begin{equation*}
    \varphi(N,\mathcal{R},E')=(b_E,E'-a_E)\ \text{if $b_E+a_E\leq E'\leq b_E+b_E$.}
\end{equation*}

The set of intervals $\{(a_E,b_E): \varphi_1(N,\mathcal{R},E)>\varphi_2(N,\mathcal{R},E)\}$ is countable because the intervals are disjoint and we can construct a one-to-one correspondence between each interval and a rational number from the interval. Hence, (\ref{description_2}) holds.

By anonymity, (\ref{description_2}) holds for each competition with two competitors. By Lemma~\ref{lemma:cons}, the description of the corresponding interval rule is the unique consistent extension of (\ref{description_2}) to competitions with more competitors.
\end{proof}
\end{theorem}

\begin{remark}
The axioms order preservation, endowment monotonicity, and consistency (used in Theorem~\ref{theorem:cons}) are logically independent.
\begin{proof}
The rule which allocates all the prize money to the competitor with the lowest position satisfies endowment monotonicity and consistency, but does not satisfy order preservation. The rule which coincides with the Equal Division rule for competitions with a prize endowment of at most one dollar, and coincides with the Winner-Takes-All rule for competitions with a higher prize endowment, satisfies order preservation and consistency, but does not satisfy endowment monotonicity. The rule from Example~\ref{ex:anonymity} satisfies order preservation and endowment monotonicity, but does not satisfy consistency.
\end{proof}
\end{remark}

\renewcommand\thelemma{\ref{lemma:homo}}
\begin{lemma}\ \\
A prize allocation rule satisfies endowment additivity if and only if it satisfies scale invariance.
\begin{proof}
Let $\varphi$ be a prize allocation rule satisfying scale invariance. Let $(N,\mathcal{R},E)$ and $(N,\mathcal{R},E')$ be two competitions and let $i\in N$ be a competitor. Then
\begin{alignat*}1
\varphi_i(N,\mathcal{R},E+E') &= (E+E')\varphi_i(N,\mathcal{R},1)
\\ &= E\varphi_i(N,\mathcal{R},1)+E'\varphi_i(N,\mathcal{R},1)
\\ &= \varphi_i(N,\mathcal{R},E)+\varphi_i(N,\mathcal{R},E').
\end{alignat*}
Hence, $\varphi$ satisfies endowment additivity.

Now, let $\varphi$ be a prize allocation rule satisfying endowment additivity. Then $\varphi$ satisfies endowment monotonicity. By Lemma~\ref{lemma:econt}, $\varphi$ satisfies endowment continuity. Let $(N,\mathcal{R},E)$ be a competition and let $i\in N$ be a competitor. If $E$ is a rational number, then there exist two natural numbers $p\in\mathbb{N}$ and $q\in\mathbb{N}$ such that $E=\frac{p}{q}$. By endowment additivity,
\begin{alignat*}1
\varphi_i(N,\mathcal{R},E) &= \varphi_i(N,\mathcal{R},\tfrac{p}{q})
\\ &= p\varphi_i(N,\mathcal{R},\tfrac{1}{q})
\\ &= \tfrac{p}{q}q\varphi_i(N,\mathcal{R},\tfrac{1}{q})
\\ &= \tfrac{p}{q}\varphi_i(N,\mathcal{R},1)
\\ &= E\varphi_i(N,\mathcal{R},1).
\end{alignat*}
By endowment continuity, $\varphi_i(N,\mathcal{R},E)=E\varphi_i(N,\mathcal{R},1)$ for each real number $E$. Hence, $\varphi$ satisfies scale invariance.
\end{proof}
\end{lemma}

\bigskip

\renewcommand\thetheorem{\ref{theorem:localcons}}
\begin{theorem}\ \\
A prize allocation rule satisfies anonymity, order preservation, winner strict endowment monotonicity, and local consistency if and only if it is a single-parametric rule.
\begin{proof}
It is readily checked that each single-parametric rule satisfies anonymity, order preservation, winner strict endowment monotonicity, and local consistency. 

Let $\varphi$ be a prize allocation rule satisfying anonymity, order preservation, winner strict endowment monotonicity, and local consistency. Then $\varphi$ satisfies endowment monotonicity. By Lemma~\ref{lemma:econt}, $\varphi$ satisfies endowment continuity. Let $N\subseteq U$ with $|N|=2$ be a set of two competitors and let $\mathcal{R}$ be the corresponding ranking. Denote $N=\{1,2\}$ such that $\mathcal{R}(1)=1$ and $\mathcal{R}(2)=2$. For each $x\in\mathbb{R}_+$, define $f(x)=y$ if and only if $\varphi_1(N,\mathcal{R},x+y)=x$ and $\varphi_2(N,\mathcal{R},x+y)=y$. By winner strict endowment monotonicity, endowment continuity, and order preservation, $f:\mathbb{R}_+\to\mathbb{R}_+$ is a well-defined, continuous and non-decreasing function with $f(x)\leq x$ for each $x$. By anonymity, $\varphi$ is a single-parametric rule with representation $f$ for each competition with two competitors. Let $\varphi'$ be a single-parametric rule with representation $f$ for each competition with arbitrary number of competitors.

Suppose for the sake of contradiction that there exists a competition $(N,\mathcal{R},E)$ such that $\varphi'(N,\mathcal{R},E)\neq\varphi(N,\mathcal{R},E)$. Denote $N=\{1,\ldots,|N|\}$ such that $\mathcal{R}(k)=k$ for each $k\in N$. Let $i\in N$ be such that $\varphi_i'(N,\mathcal{R},E)\neq\varphi_i(N,\mathcal{R},E)$ and $\varphi_j'(N,\mathcal{R},E)=\varphi_j(N,\mathcal{R},E)$ for each $j\in N$ with $j<i$. Suppose without loss of generality that $\varphi_i'(N,\mathcal{R},E)>\varphi_i(N,\mathcal{R},E)$. By local consistency and anonymity,
\begin{alignat*}1
\varphi_{i+1}'(N,\mathcal{R},E) &= \varphi_{i+1}'(\{i,i+1\},\mathcal{R}_{\{i,i+1\}},\varphi_i'(N,\mathcal{R},E)+\varphi_{i+1}'(N,\mathcal{R},E))
\\ &= f(\varphi_i'(N,\mathcal{R},E))
\\ &\geq f(\varphi_i(N,\mathcal{R},E))
\\ &= \varphi_{i+1}(\{i,i+1\},\mathcal{R}_{\{i,i+1\}},\varphi_i(N,\mathcal{R},E)+\varphi_{i+1}(N,\mathcal{R},E))
\\ &= \varphi_{i+1}(N,\mathcal{R},E).
\end{alignat*}
In a similar way, this implies that $\varphi_{i+2}'(N,\mathcal{R},E)\geq\varphi_{i+2}(N,\mathcal{R},E)$. Continuing this reasoning, $\varphi_j'(N,\mathcal{R},E)\geq\varphi_j(N,\mathcal{R},E)$ for each $j\in N$ with $j>i$. This means that
\begin{equation*}
    \sum_{j\in N}\varphi_j'(N,\mathcal{R},E)>\sum_{j\in N}\varphi_j(N,\mathcal{R},E)=E.
\end{equation*}
This is a contradiction. Hence, $\varphi'(N,\mathcal{R},E)=\varphi(N,\mathcal{R},E)$ for each competition $(N,\mathcal{R},E)$.
\end{proof}
\end{theorem}

\begin{remark}
The axioms anonymity, order preservation, winner strict endowment monotonicity, and local consistency (used in Theorem~\ref{theorem:localcons}) are logically independent.
\begin{proof}
The rule from Example~\ref{ex:anonymity} satisfies order preservation, winner strict endowment monotonicity, and local consistency, but does not satisfy anonymity. The geometric rule with representation $\lambda=2$ satisfies anonymity, winner strict endowment monotonicity, and local consistency, but does not satisfy order preservation. The rule from Example~\ref{ex:step} satisfies anonymity, order preservation, and local consistency, but does not satisfy winner strict endowment monotonicity. The rule which coincides with the ED rule for competitions with two competitors and coincides with the WTA rule for competitions with more than two competitors satisfies anonymity, order preservation, and winner strict endowment monotonicity, but does not satisfy local consistency.
\end{proof}
\end{remark}

\bigskip

\renewcommand\thetheorem{\ref{theorem:topcons}}
\begin{theorem}\ \\
A prize allocation rule satisfies anonymity, order preservation, winner strict endowment monotonicity, and top consistency if and only if it is a parametric rule.
\begin{proof}
It is readily checked that each parametric rule satisfies anonymity, order preservation, winner strict endowment monotonicity, and top consistency. 

Let $\varphi$ be a prize allocation rule satisfying anonymity, order preservation, winner strict endowment monotonicity, and top consistency. Then $\varphi$ satisfies endowment monotonicity. By Lemma~\ref{lemma:econt}, $\varphi$ satisfies endowment continuity. We show that $\varphi$ is a parametric rule by induction on the number of competitors. Clearly, $\varphi_i(N,\mathcal{R},E)=E=f_1(x)$ holds for each competition $(N,\mathcal{R},E)$ with $|N|=1$ and each competitor $i\in N$. Let $n\in\mathbb{N}$ and assume that there exist continuous and non-decreasing functions $f_1,f_2,\ldots,f_n:\mathbb{R}_+\to\mathbb{R}_+$ with $f_1(x)=x$ for each $x$ and $f_{k+1}(x)\leq f_k(x)$ for each $k$ such that for each competition $(N,\mathcal{R},E)$ with $|N|\leq n$ and each competitor $i\in N$, we have $\varphi_i(N,\mathcal{R},E)=f_{\mathcal{R}(i)}(x)$, where $x\in\mathbb{R}_+$ is such that $\sum_{k=1}^{|N|}f_k(x)=E$. 

Let $N=\{1,\ldots,n+1\}$ be a set of $n+1$ competitors and let $\mathcal{R}$ be the corresponding ranking such that $\mathcal{R}(i)=i$ for all $i\in N$. By order preservation, endowment continuity, and winner strict endowment monotonicity, $\varphi_1(N,\mathcal{R},E)$ is unbounded, continuous and increasing function of~$E$. Hence, for each $x\in\mathbb{R}_+$, there is a unique $E$ such that $\varphi_1(N,\mathcal{R},E)=x$, we can define $f_{n+1}(x)=\varphi_{n+1}(N,\mathcal{R},E)$ and, by top consistency and inductive hypotheses, $\varphi_2(N,\mathcal{R},E)=f_2(x),\ldots,\varphi_{n}(N,\mathcal{R},E)=f_n(x)$. By endowment continuity, winner strict endowment monotonicity, and order preservation, $f_{n+1}:\mathbb{R}_+\to\mathbb{R}_+$ is a continuous and non-decreasing function with $f_{n+1}(x)\leq f_n(x)$ for each $x$. By anonymity, $\varphi$ is a parametric rule with representation $f_1,f_2,\ldots,f_{n+1}$ for each competition with $n+1$ competitors.
\end{proof}
\end{theorem}

\begin{remark}
The axioms anonymity, order preservation, winner strict endowment monotonicity, and top consistency (used in Theorem~\ref{theorem:topcons}) are logically independent.
\begin{proof}
The rule from Example~\ref{ex:anonymity} satisfies order preservation, winner strict endowment monotonicity, and top consistency, but does not satisfy anonymity. The geometric rule with representation $\lambda=2$ satisfies anonymity, winner strict endowment monotonicity, and top consistency, but does not satisfy order preservation. The rule from Example~\ref{ex:step} satisfies anonymity, order preservation, and top consistency, but does not satisfy winner strict endowment monotonicity. The rule which coincides with the ED rule for competitions with two competitors and coincides with the WTA rule for competitions with more than two competitors satisfies anonymity, order preservation, and winner strict endowment monotonicity, but does not satisfy top consistency.
\end{proof}
\end{remark}

\bibliographystyle{apalike}
\bibliography{race}

\end{document}